\documentclass{aa}
\usepackage{natbib,twoopt}
\usepackage{graphicx}
\usepackage{titlesec} \usepackage{amsmath, amssymb}
\usepackage{tikz}
\usepackage{float}
\usepackage[inline]{enumitem}
\usetikzlibrary{positioning, arrows, shapes, calc}
\usetikzlibrary{shapes.multipart}
\usepackage{xcolor}
\usepackage{txfonts}
\usepackage{comment}
\usepackage{longtable}
\usepackage{ulem}
\usepackage{comment}
\usepackage{appendix}
\usepackage{pgfplotstable}
\usepackage{placeins}
\usepackage{mathtools}
\usepackage[T1]{fontenc}
\usepackage{csvsimple}
\usepackage{booktabs} 

\usepackage{xltabular}
\usepackage{booktabs}
\usepackage{hyperref}
\usepackage{tabularx}

\newcommand{\hiprepI}{$\eta$~Boo}
\newcommand{\typerepI}{F~dwarf}
\newcommand{\classrepI}{p03\_F-dw}

\newcommand{\hiprepII}{HD~49933}
\newcommand{\typerepII}{F~dwarf}
\newcommand{\classrepII}{m04\_F-dw}

\newcommand{\hiprepIII}{18 Sco}
\newcommand{\typerepIII}{G~dwarf}
\newcommand{\classrepIII}{00\_G-dw}

\newcommand{\hiprepIV}{HD~140283}
\newcommand{\typerepIV}{G~subgiant}
\newcommand{\classrepIV}{m24\_G-sg}

\newcommand{\hiprepV}{61~Cyg~A}
\newcommand{\typerepV}{K~dwarf}
\newcommand{\classrepV}{m03\_K-dw}

\newcommand{\hiprepVI}{HD~182736}
\newcommand{\typerepVI}{G~subgiant}
\newcommand{\classrepVI}{m02\_G-sg}

\newcommand{\hiprepVII}{$\beta$~Gem}
\newcommand{\typerepVII}{Clump giant}
\newcommand{\classrepVII}{m01\_Cl-G}

\newcommand{\hiprepVIII}{HD~175305}
\newcommand{\typerepVIII}{Clump giant}
\newcommand{\classrepVIII}{m15\_Cl-G}

\newcommand{\hiprepIX}{Arcturus}
\newcommand{\typerepIX}{Bright giant}
\newcommand{\classrepIX}{m05\_Br-G}

\newcommand{\hiprepX}{HD~122563}
\newcommand{\typerepX}{Bright giant}
\newcommand{\classrepX}{m27\_Br-G}

\begin{document}

   \title{Gaia FGK benchmark stars:}
   \subtitle{Abundances of \textit{n}-capture elements of the third version}
\titlerunning{}
\author{
    S. Vitali\inst{1,2} \and P. Jofré\inst{3} \and L. Casamiquela\inst{4}\and C. Soubiran\inst{5} \and U. Heiter\inst{6} \and
    C. Aguilera-Gómez\inst{7} \and D. Barrios-L\'opez\inst{3} \and  S. Blanco-Cuaresma\inst{8,9} \and A. Escorza\inst{1,2} \and I. Hernández–Araya\inst{7} \and T. Signor\inst{3,10} \and H. Sinclair-Wentworth\inst{11} \and C. Worley\inst{11}  
    }
    
    \institute{Instituto de Astrofísica de Canarias, C. Vía Láctea, s/n, 38205 La Laguna, Santa Cruz de Tenerife, Spain.\\
    \email{svitali@iac.es}
      \and
    Universidad de La Laguna, Dpto. Astrofísica, Av. Astrofísico Francisco Sánchez, 38206 La Laguna, Santa Cruz de Tenerife, Spain.
      \and
    Instituto de Estudios Astrofísicos, Facultad de Ingeniería y Ciencias, Universidad Diego Portales, Av. Ejército Libertador 441, Santiago, Chile.
      \and
    LIRA, Observatoire de Paris, Université PSL, Sorbonne Université, Université Paris Cité, CY Cergy Paris Université, CNRS, 92190 Meudon, France.
      \and
    Laboratoire d’Astrophysique de Bordeaux, Univ. Bordeaux, CNRS, B18N, allée Geoffroy Saint-Hilaire, 33615 Pessac, France.
      \and
    Observational Astrophysics, Department of Physics and Astronomy, Uppsala University, Box 516, SE-751 20 Uppsala, Sweden.
       \and 
    Instituto de Astrofísica, Pontificia Universidad Católica de Chile, Av. Vicuña Mackenna 4860, 782-0436 Macul, Santiago, Chile. 
          \and 
    Harvard-Smithsonian Center for Astrophysics, 60 Garden Street, Cambridge, MA 02138, USA.
       \and 
    Faculty of Psychology, UniDistance Suisse, Brig, Switzerland.
       \and 
    Inria Chile Research Center, Av. Apoquindo 2827, piso 12, Las Condes, Santiago, Chile.
      \and
    School of Physical and Chemical Sciences - Te Kura Matū, University of Canterbury, New Zealand.}

 \date{Received \today, accepted }

  \abstract
   {In the current era, in which an unprecedented wealth of data are available for the study of the Milky Way, Gaia benchmark stars (GBSs) have become an established reference and calibration sample. Studies of stellar structure and evolution and of the chemical history of our Galaxy generally rely on large spectroscopic surveys and their output catalogs. In this context, deriving precise and accurate stellar parameters and chemical abundances is of paramount importance.} 
  {This study provides the determination of neutron-capture element abundances and extends the set of chemical abundances available for the third GBS release (GBSv3).}
  {Based on the compilation of high-resolution spectra assembled for GBSv3 and consistent with the spectral analysis adopted for the chemical abundances of GBSv3, we used the public iSpec code to derive heavy element abundances.}
  {We inferred homogeneous abundances of neutron-capture elements (Y, Zr, Mo, Ba, La, Ce, Pr, Nd, and Eu) across the GBSv3 sample using an in-depth line assessment tailored to different groups identified through a clustering algorithm that accounts for the diversity in stellar parameters and metallicities. This approach addresses key challenges in the spectral analysis of these elements, including the paucity of usable lines, weak line strengths, saturation effects, and sensitivity to atomic data. The assessment yielded reliable measurements, establishing an extended and robust reference scale in good agreement with the literature.}
  {This compilation of neutron-capture abundances is based on the GBS sample's robust and accurate atmospheric parameters and the analysis of a large sample of stellar spectra per star, which provides a reliable and homogeneous spectral analysis. It also supports the use of chemical abundances as precise tracers of the Milky Way’s star formation history and chemical evolution and constitutes a legacy sample for the calibration of current and future spectroscopic surveys.}
  {}


   \maketitle

\section{Introduction}

In modern astronomy, the amount of data produced by large spectroscopic surveys aimed at studying the history and evolution of the Galaxy through stellar abundances has reached an unprecedented level. As more data releases become available (e.g., APOGEE \citealt{2022ApJS..259...35A}, GALAH \citealt{2025Buder}, SDSSV \citealt{2026Kollmeier}, 4MOST \citealt{2019deJong}, WEAVE \citealt{2024Jin}), clearly disentangling genuine astrophysical diversity from the differences introduced by methodology may become difficult. In this context, the astronomical community is increasingly recognizing the need for robust and reliable calibrators to homogenize the outputs of different catalogs, which can be affected by different internal analysis methods and calibration strategies. 

Beyond the more “classical” spectroscopic efforts, various machine learning approaches have been developed to ensure consistency across different data releases \citep[e.g.,][]{2015Ness,2017Ho,2025Guiglion}. Despite these advances, the long history of development and use of the Gaia FGK benchmark stars (GBSs) demonstrates that validation samples grounded in classical spectroscopy remain essential. Their assembly began over a decade ago with \citet[][hereafter PI]{2015heiter}, who presented the fundamental temperatures and surface gravities for the first sample. \citet[][PII]{2014blanco-cuaresma} introduced the first spectral library of these stars, which was analyzed homogeneously to derive metallicities in \citet[][PIII]{2014jofre} and abundances of $\alpha$- and iron-peak elements in \citet[][PIV]{2015jofre}. It was soon recognized that the original sample lacked metal-poor stars, prompting \citet[][PV]{2016hawkins} to propose additional metal-poor candidates. Because each star in the sample is distinct, the GBSs provide an excellent resource for investigating various aspects of spectroscopy. Systematic uncertainties associated with methods for deriving abundances were explored in \citet[][PVI]{2017jofre}, and other studies have used these stars to test methods or instruments \citep[e.g.,][]{2018Buder,2020A&A...642A.182A,2022Gent,2025Buder}. 

With new data from Gaia and improved angular diameters, the GBS sample was expanded to 200 stars by \citet[][PVII]{2024soubiran}. A spectral library and homogeneous estimates of metallicities, $\alpha$-capture, and Fe-peak elements were subsequently presented in \citet[][PVIII]{2025Casamiquela}. 

In this work, we build on these efforts to extend PVIII by providing a uniform set of neutron-capture (n-capture) element abundances. These heavy elements (beyond the iron peak, Z $\gtrsim 30$) are produced through processes that occur during stellar evolution or in explosive astrophysical events \citep{2004Travaglio,2011RvMP...83..157K,2018Spite,2021Cowan}. Much remains to be understood about the specific astrophysical sources responsible for their synthesis \citep[see a recent review by][]{2026Thielemann}. Precise measurements of this family of elements are therefore of paramount importance, as they are widely used in Galactic studies, providing strong constraints on the star formation history of a stellar population \citep{2019Hansen,2023moler,2026Anoardo}. The need to better constrain these elements is also reflected in the design of future observing facilities \citep[e.g.,][]{2020Msngr.180...10C,2024Mainieri,2023Magrini}, which will cover spectral wavelength regions specifically selected to maximize the information obtained on n-capture species.

As the latest contribution to the ongoing series of GBS papers mentioned above, this work presents key results that serve as a cornerstone for the calibration and validation of ongoing and future efforts in heavy element abundance measurements. It is structured as follows. Sect.\ref{sect:2} provides an overview of the adopted analysis strategy. Sect.~\ref{sect:3} presents the results for ten representative GBSs selected to define different line selections according to stellar type. The resulting abundances are assessed through comparisons with values from the literature. In Sect.~\ref{sect:4}, the previous line selection is validated, and the analysis is extended to the full GBS version 3 (GBSv3) sample. A discussion focusing on the solar case and the impact of these results on Galactic studies is also presented. Finally, in Sect.~\ref{sect:5}, we summarize our main findings and provide details on data accessibility.

\section{Analysis strategy}\label{sect:2}
This work is a continuation of PVIII. Therefore, the data description and details of the spectroscopic analysis can be found in that paper. We briefly summarize the methodology in the following sections and present our strategy to estimate n-capture abundances for the GBSv3.

\subsection{Spectroscopic analysis}
The pipeline adopted in PVIII started with spectral pre-processing. The spectra are from a collection of public archive data and our own observations, and for most of the stars spectra obtained with more than one instrument were available. A homogeneous dataset was assembled, meaning that the spectra were placed on the same wavelength range (480 to 680~nm), co-added for each star to increase the final signal-to-noise ratio (S/N) per instrument, normalized, convolved to a common resolution, corrected for radial velocity, and resampled. All these steps were performed using the functionalities of \texttt{iSpec} \citep[version 2020.10.01;][]{2014BlancoB,2019BlancoCuaresma}. 

As described in PVIII, abundances were derived using {\tt iSpec} by on-the-fly spectral synthesis using the sixth version of the \textit{Gaia}–ESO (GES) line list \citep{Heiter21}, enriched with molecular data \citep{2023Gerber}. The atmospheric models are taken from MARCS \citep{2008Gustafsson}, and solar abundances from \citep{grevesse07}. This implies that the reported abundances are derived under the assumptions of local thermodynamic equilibrium (LTE) and 1D modeling. Abundances were derived using the stellar parameters from PVII, which were determined from fundamental methods as described in that work. In PVIII only the 
$\alpha$ and Fe-peak elements were published, although the analysis was initiated for a wider set of species. Here, we extend it to the n-capture elements (Y, Zr, Mo, Ba, La, Ce, Pr, Nd, Eu) performing additional line selection and validation of the spectral fits using \texttt{iSpec}. Although outputs from different radiative transfer codes are discussed in PVIII, here we adopt the results obtained with Turbospectrum \citep{1998Alvarez,2012Plez,2023Gerber}. As detailed in PVIII, this code was selected to recommend the final abundances of Fe-peak and $\alpha$ elements because it is optimized to handle a large number of molecular transitions, a feature that is particularly important for cool stars ($T_{\mathrm{eff}} < 4500$ K). In PIV and PVIII, most elements had more than ten lines (except Mg), allowing for a line selection based on line-to-line scatter and consistency with the solar spectrum. In contrast, as described in this work, a more detailed evaluation and assessment is required for each element. 

To derive the final abundances, we adopted a two-step approach. First, we selected ten stellar groups each represented by a single star (as described in Sect.~\ref{sect:repgbs}), which we call the representative star, or RepGBS. We carefully evaluated the lines suitable for each element and RepGBS. We then determined the optimal set of heavy element lines for each stellar group and applied this selection to all the stars in the group associated with the specific RepGBS. The nomenclature of the groups, along with the clustering algorithm performed directly on the spectra and used to assign stars to each group, is presented in the next section.

\subsection{Stellar grouping based on representative stars}\label{sect:repgbs}
\begin{figure}[t!]
    \centering    \includegraphics[width=0.49\textwidth]{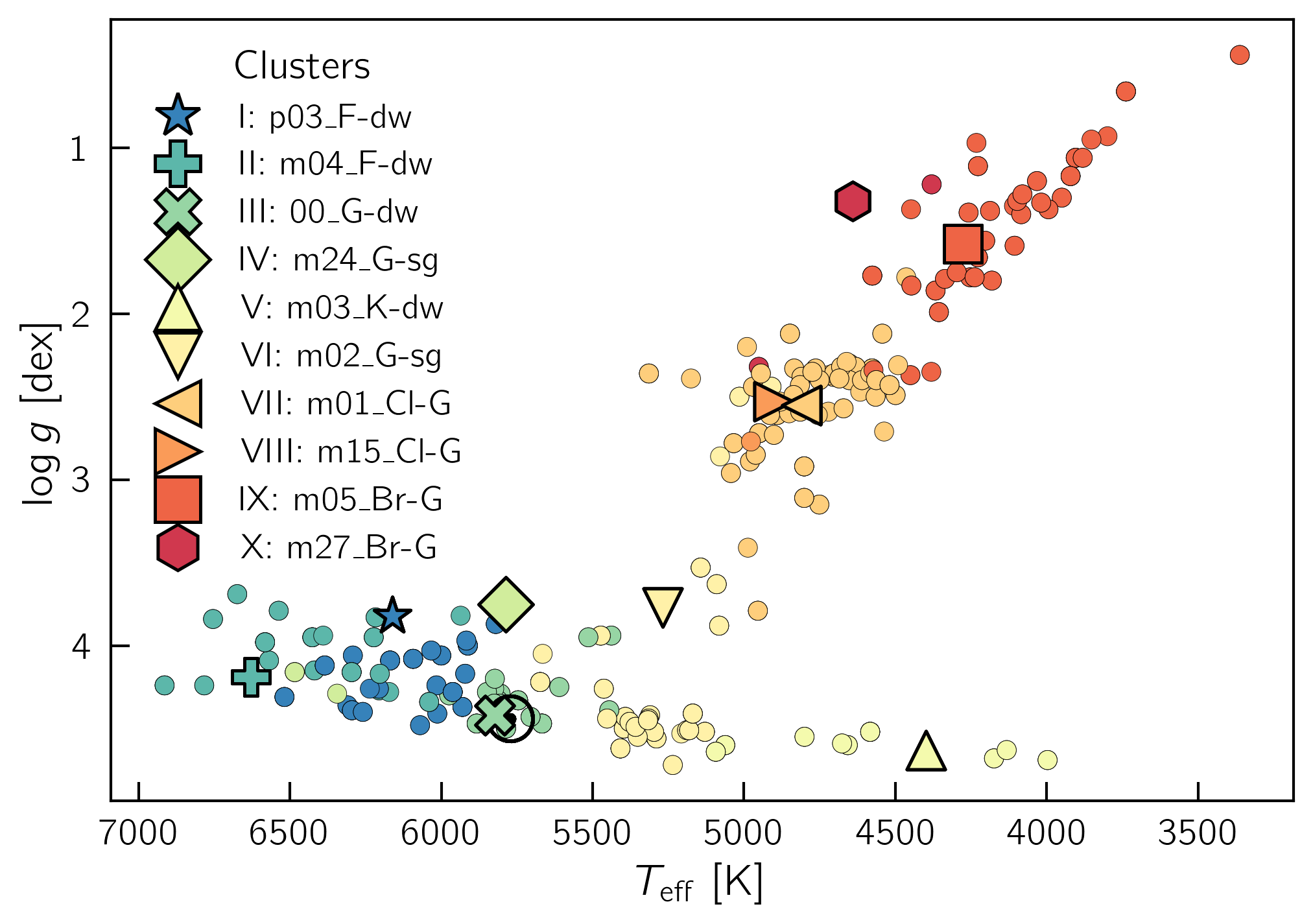}    \includegraphics[width=0.49\textwidth]{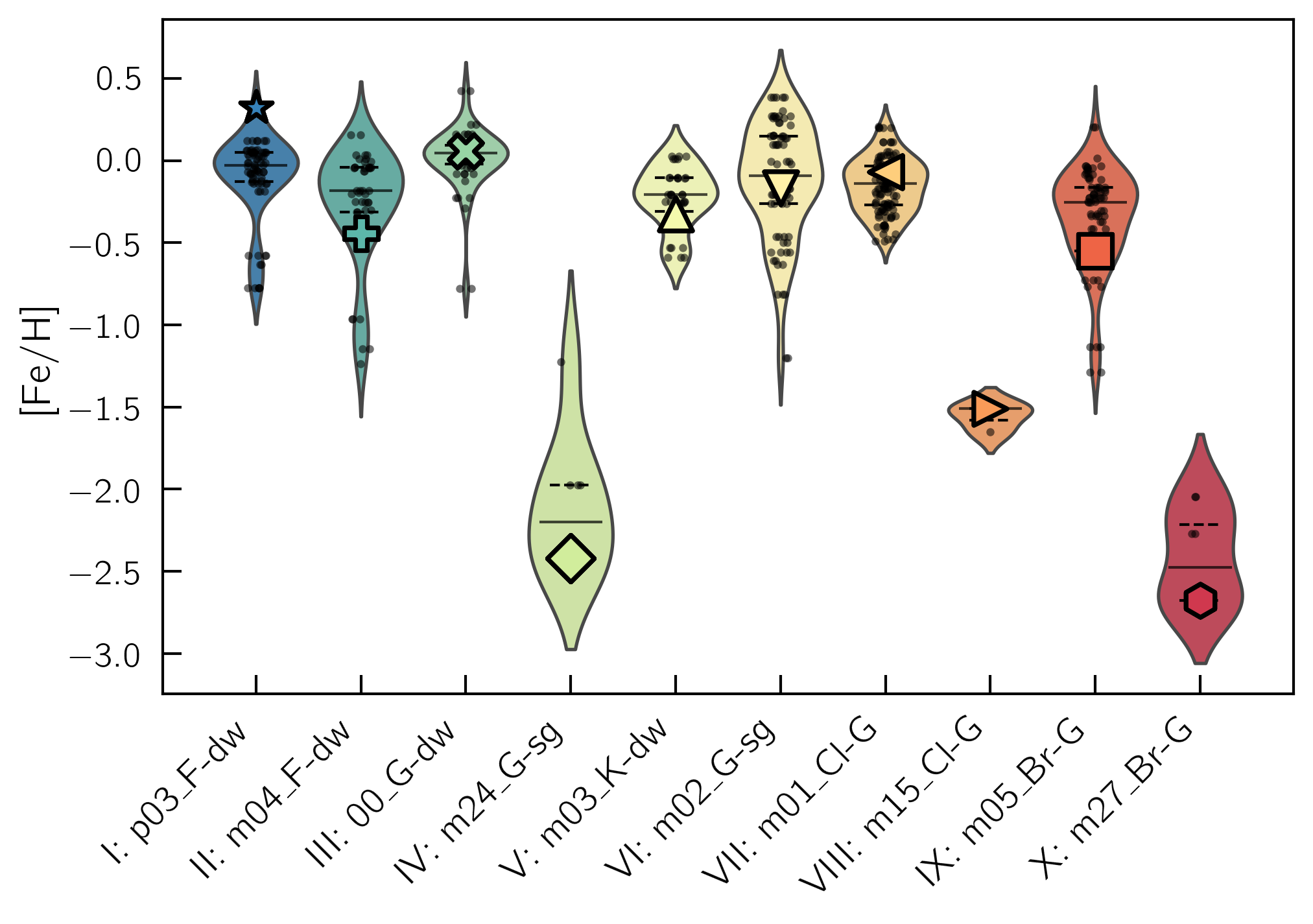}  
    \caption{Kiel diagram and parameter distributions of the stellar sample, color coded by stellar group. \textbf{Top:} Kiel diagram where normal stars are shown as circles, while the RepGBSs are represented with distinct symbols. \textbf{Bottom:} Violin plot of the metallicity for each group. The width of each violin reflects the density of stars, with inner lines indicating the median and quartiles. Individual stars are overplotted in gray, with the RepGBS shown using the same symbols as in the Kiel diagram.}
    \label{fig:hr_kmeans}
\end{figure}

To assess the selection of stars in each group, we used the ten representative stars adopted in PVIII (see their Table~1) as RepGBS. They span the parameter space and [Fe/H] range of the GBS sample. These RepGBSs are represented as ten distinct symbols, and their group nomenclature is reported both in Table~\ref{tab:repgb_nomenclature} and in the legend of the  Kiel diagram at the top of Fig.~\ref{fig:hr_kmeans}. We emphasize that more refined classifications could likely be made, and the current selection may not necessarily be the best representation of the GBSv3 parameter space. Nevertheless, we retained these ten groups, as the RepGBSs in each group have been observed with multiple spectra, allowing for a robust evaluation of the line-fitting quality, a key step in our analysis.

We defined stellar groups with the aim that, within each class, a common line list could be used to derive chemical abundances. This requirement is primarily governed by the atmospheric parameters $T_{\rm eff}$, $\log g$, and [Fe/H], which determine the strength and sensitivity of spectral lines. We recast this problem as a projection task: We sought a transformation of the spectra into a low-dimensional space in which stars with similar atmospheric parameters are clustered around a set of predefined representative stars. Each group is therefore anchored to one representative star, and all other stars are assigned to the closest representative in this transformed space.

To avoid instrumental effects that could bias the classification, we restricted the analysis to some spectral regions that are sensitive to the stellar atmospheric parameters $T_{\rm eff}$, $\log g$, and [Fe/H].
Specifically, we include the Balmer lines H$\beta$ ($485.8-486.5$ nm) and H$\alpha$ ($656.0-656.6$ nm), the \ion{Mg}{I} b triplet ($516.5-518.5$ nm), and the \ion{Na}{I} D doublet ($588.8-589.8$ nm), together with a set of selected \ion{Fe}{I} lines spanning the $487-559$ nm range. We then removed, for all spectra, any pixels with missing flux in any spectrum. This resulted in a total of 4324 pixels.

This procedure was carried out in two steps. First, we constructed a linear discriminant analysis (LDA; \citealt{hastie2009elements}) model using the spectra of the ten RepGBS (a total of 40 spectra). The LDA finds a linear projection that maximizes the separation between these predefined classes while minimizing the intra-class variance. We restricted the representation to three dimensions in order to capture the dominant discriminative directions while avoiding poorly constrained directions given the limited number of available spectra. The learned transformation was then applied to the full stellar sample. Finally, each star was assigned to the group of the closest RepGBS using a Euclidean distance metric in the LDA space. In practice, this is equivalent to a nearest-centroid classification, which we implemented using a k-means algorithm (\citealt{lloyd1982least}) initialized on the RepGBS.

From the Kiel diagram in Fig.~\ref{fig:hr_kmeans}, we observed that the resulting clustering does not produce cubical groups in stellar parameters but rather stripes across the  diagram. This reflects the interdependency of stellar parameters and their combined effect on the overall shape of stellar spectra. The bottom panel of the figure shows the [Fe/H] distribution for each group as violin plots. These clearly show a wider spread for the metal-poor groups IV and X, indicating that at low metallicity the differences between stars are smaller, making it more difficult to discriminate between distinct groups. This increasing difficulty in identifying discriminating features in lower-metallicity spectra could be mitigated by including additional representative stars. The limitations and implications of this choice in our final results are discussed later in the text (Sect.~\ref{sect:val_clust}). 

Two stars (HIP~47908 and HIP~98269) become part of Group~I, \classrepI,  in our algorithm. We recall that our clustering algorithm is based on distance metrics which are calculated from a selection of spectral regions only (see above).  The distribution of the parameters in the Kiel diagram and metallicity distribution in Fig.~\ref{fig:hr_kmeans} serves as a validation diagnostics but are not used in the clustering. Since the location in the Kiel diagram of these two stars is closer to the stars of Group~VII and VIII, we manually assigned them to Group~VII (Group~VIII is significantly more metal-poor than HIP~47908 and HIP~98269).

\begin{figure}[t]
    \centering
    \includegraphics[width=0.99\linewidth]{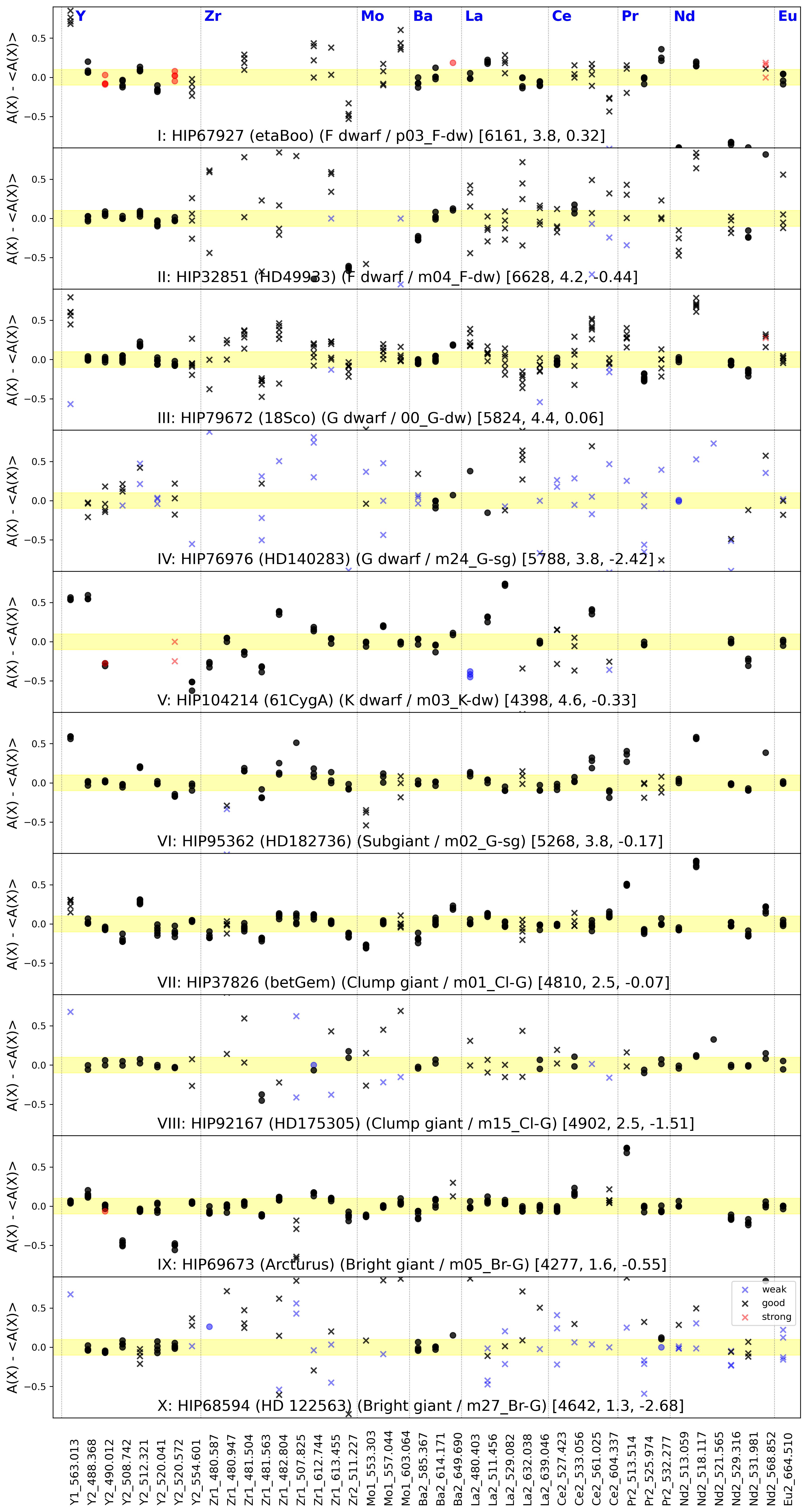}
    \caption{Compilation of spectral lines for each n-capture element analyzed in this work, based on an analysis of the RepGBS. The vertical axes show the differences between the abundances measured for each line and each instrument and the median of all measurements per element. Crosses indicate lines with instrument-to-instrument dispersion above 0.15 dex, while circles represent lines with lower dispersion. The yellow band is centered on zero and is 0.2 dex wide. Red symbols indicate strong lines with REW>-4.5, blue symbols are weak lines with REW<-6.7, and lines with black symbols fall in between, thus fulfilling our REW criterion (see text). The labels on the horizontal axis specify the element symbol, ionization stage (1 = neutral, 2 = ionized), and central wavelength in nm}
    \label{fig:lines_new}
\end{figure}

\subsection{Line selection}\label{sect:line_sel}

We began our selection by considering lines for which quality flags are provided in the GES line list study \citep{Heiter21}.  As described in PVIII, each line is associated with two flags, \texttt{gfflag} and \texttt{synflag}, with possible values of “yes” (\texttt{Y}), “undetermined” (\texttt{U}), and “no” (\texttt{N}). These flags respectively indicate the reliability of the atomic $\log gf$ value and the degree of blending from neighboring lines. However, given the paucity of lines per element, the PVIII criterion, requiring  \texttt{gfflag} and \texttt{synflag} flagged as positive in the GES list, is no longer applicable. Therefore in some cases we complemented the selection with additional lines (e.g., \ion{La}{II} and \ion{Mo}{I}). The selection of the optimal set of lines is instead based on a careful assessment of each fit, guided by the \texttt{iSpec} diagnostics (e.g., $\chi^2$, abundance uncertainties, and  equivalent width values), together with a thorough comparison with the literature (Sect.~\ref{sect:literature_comp}). 

Our goal was twofold: preserve measurement accuracy by rejecting lines that show excessive scatter and retain as many measurements as possible (i.e., multiple lines across different spectra) to ensure robust statistical precision. In this context, we inspected the line-by-line abundances for each RepGBS and identified a set of lines from which to derive the final abundances, tailored to each group presented in Sect.~\ref{sect:repgbs}, while taking into account the differences in evolutionary phase and, consequently, in atmospheric parameters among the GBSv3.

Figure~\ref{fig:lines_new} shows the abundance measurements obtained with {\tt iSpec} for all the lines included in our analysis for the ten RepGBSs in different panels. The presence of multiple measurements per line and per star arises from the fact that the RepGBS have several spectra obtained with different spectrographs: HARPS \citep{2003Mayor}, UVES \citep{2000Dekker}, FEROS \citep{1999Msngr..95....8K}, NARVAL \citep{2003EAS.....9..105A}, ESPaDOnS \citep{Donati2006}, and ELODIE \citep{1996Baranne} (for more details on the instruments and data reduction, see Sect.~2.1 of PVIII).

Since our \texttt{iSpec} analysis also provides equivalent widths (EWs), it is possible to assess line strength, which is indicated by the color of the symbols. Following PVIII, we apply a cut on reduced EW (REW\footnote{ $\mathrm{REW} = \log_{10}(\mathrm{EW}/\lambda)$}) range, namely $-6.7 < \mathrm{REW} < -4.5$.  

From Fig.~\ref{fig:lines_new} we can see that the number of lines per element varies from one to nine. When no result is provided, it means that the pipeline was unable to fit the feature, most likely due to excessive weakness, a gap in the spectrum, or a high noise level. The diversity in stellar parameters leads to different symbols and colors in this figure, which reflect the wide range of line strengths, and line-by-line variations from star to star. This suggests that we have different numbers of usable lines per star and thus a single selection criterion for all stars cannot be applied. 

As a starting point, we therefore consider only measurements flagged with circles. Subsequently, we inspect the abundances on a line-by-line basis, since some elements rely on very weak lines (e.g., \ion{Zr}{I} or \ion{Ce}{II}), and some transitions yield systematically discrepant abundances despite appearing reliable (e.g., \ion{Nd}{II}). For the final abundances, saturated lines (in red) are generally rejected while weak lines (in blue) are instead retained when the fit quality is satisfactory, typically indicated by low spectrum-to-spectrum scatter. The decision to reject or retain a measurement was made through visual inspection of all lines in Fig.~\ref{fig:lines_new}.  

\section{Neutron-capture abundances of the RepGBS}\label{sect:3}

In this section, we discuss each element to derive the final chemical abundances for the RepGBS. We first explain how we computed the final abundance and then discuss each element in detail. 

\subsection{Final abundances}\label{sect:3.1}

Once the lines were selected, 
we computed the final abundance $\overline{A}(x)$ using a unified weighted approach that combines all available measurements:

\begin{equation} \label{eq:abund}
\overline{A}(x) = \frac{\sum_{i=1}^{N_{\text{spec}}} \sum_{j=1}^{N_{\text{lines}}^{(i)}} w_{ij} A_{ij}(x)}{\sum_{i=1}^{N_{\text{spec}}} \sum_{j=1}^{N_{\text{lines}}^{(i)}} w_{ij}},
w_{ij} = \frac{1}{\sigma_{ij}^2}.
\end{equation}

\noindent Here, $A_{ij}(x)$ is the abundance from line $j$ in spectrum $i$, $\sigma_{ij}$ is the uncertainty on $A_{ij}(x)$, $N_{\text{spec}}$ is the number of spectra for the star, and $N_{\text{lines}}^{(i)}$ is the number of lines in spectrum $i$. 

This approach seamlessly handles both single- and multiple-spectrum cases within a unified framework, naturally accounting for measurement precision and line-to-line scatter to provide robust abundance estimates. The ten RepGBSs, in particular, have multiple spectra taken from different spectrographs, which ensures that the lines passing our selection are reliable. Our line selection thus allowed us to confidently derive abundances in the 31 cases where only one spectrum or one line per star is available.

Following Paper~VIII, we computed the unbiased weighted standard deviation by adopting their Eq.~4: 
\begin{equation}\label{uncert1}
\sigma =
\sqrt{
\frac{
\sum_{i=1}^{N_{\mathrm{spec}}} \sum_{j=1}^{N_{\mathrm{lines}}^{(i)}} 
w_{ij} \left(A_{ij}(x) - \overline{A}(x)\right)^2
}{
\sum_{i=1}^{N_{\mathrm{spec}}} \sum_{j=1}^{N_{\mathrm{lines}}^{(i)}} w_{ij} 
- \frac{ \sum_{i=1}^{N_{\mathrm{spec}}} \sum_{j=1}^{N_{\mathrm{lines}}^{(i)}} w_{ij}^2 }
       { \sum_{i=1}^{N_{\mathrm{spec}}} \sum_{j=1}^{N_{\mathrm{lines}}^{(i)}} w_{ij} },
}
}
\end{equation}

\noindent where $w_{ij} = 1/\sigma_{ij}^2$ are the weights associated with each line measurement, and $\overline{A}(x)$ is the weighted mean abundance (Eq.~\ref{eq:abund}). This quantity characterizes the internal consistency of the abundances derived from different spectral lines, while accounting for the individual measurement uncertainties $\sigma_{ij}$, which encode the quality of the line fits, including spectral quality and line blending effects. When only a single line is available, the uncertainty is taken as the error of the line fit.

\subsection{Individual abundances}\label{sect:abundances}

Table~\ref{tab:lines} lists, for each RepGBS, the lines retained in at least one spectrum (indicated with \textcolor{green}{$\boldsymbol{\checkmark}$}) to estimate the final abundance. We note that, even when a line is selected, measurements from lower-quality spectra, affected by continuum placement, normalization issues, reduction artifacts, or low S/N, are excluded from the abundance determination. The resulting abundances and uncertainties (Eqs.~\ref{eq:abund}, \ref{uncert1}) are shown in the figures of this section and provided as online material for the full GBSv3 sample. Elements with no reliable abundance, due to problematic line fits or unreliable features, are reported in Appendix~\ref{appendix:other}.

\subsubsection{Yttrium}
For yttrium (Y), several absorption lines are available. However, abundances derived from \ion{Y}{I} lines were discarded, as these features are weak, frequently blended, and sensitive to non-LTE effects \citep{2023Alexeeva}. As shown in Fig.~\ref{fig:lines_new}, more than half of the cases are marked with crosses, indicating lines that are either too weak or strongly discrepant with the \ion{Y}{II} measurements. Even when \ion{Y}{I} lines are plotted with black circles (e.g., panels V and VI), their abundances remain systematically higher than those from \ion{Y}{II}, suggesting unresolved blends. This effect is more pronounced in panel V (K dwarf) and panel IX (bright giant), where blending is particularly severe.

A few lines in Panel~I are marked in red, namely Y2\_490.021 and Y2\_520.572, indicating that they do not satisfy the REW criterion and may be affected by saturation; only three lines are considered suitable. Mild saturation is also present for Y2\_490.021 in \hiprepIX, the coolest star in the sample, where the line falls in a region of continuum suppression. The \ion{Y}{II} line at 520.572~nm lies on the wings of a strong \ion{Fe}{I} feature, making it difficult to measure and leading to a lower derived abundance in this star. A similarly low abundance was obtained for Y2\_508.742. For both lines, this behavior is likely due to a combination of blending and continuum suppression in this spectral region in cooler stars such as \hiprepIX. This effect may be further enhanced by molecular absorption (e.g., TiO bands around 500~nm; \citealt{2019McKemmish}), which is not fully captured in the synthesis, although disentangling it from general line crowding is not straightforward. These lines, however, were used for \classrepVII\ \hiprepVII, yielding results consistent with the median of the other lines.

The Y2\_554.601 line was kept for \hiprepIX\ and for the more metal-rich star \hiprepI\ and the two slightly subsolar metallicity stars \hiprepVII\ and \hiprepVI. However, it was discarded in the remaining dwarfs and in the metal-poor giants because its weakness prevents the synthesis from properly resolving the blends.
\begin{figure}[t]
    \centering   \includegraphics[width=0.99\linewidth]{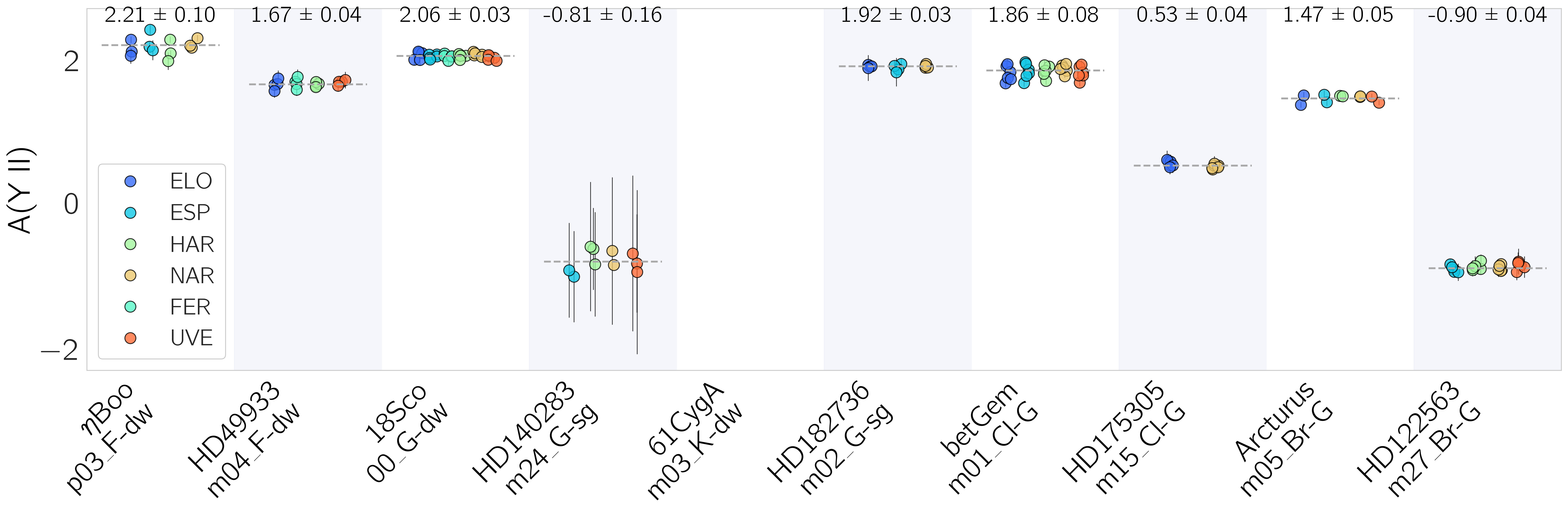}
    \caption{Y abundances derived from the selected lines for each star. Panels have been left empty when no measurable lines were available. Each panel reports the final mean abundance, $\overline{A}(X)$ (Eq.~\ref{eq:abund}), computed from the lines listed in Table~\ref{tab:lines} along with the weighted standard deviation, $\sigma$. Individual measurements from the different spectrographs are shown as colored points, while the dashed line indicates the final mean abundance.}  \label{fig:yII}
\end{figure}
The Y2\_512.321 line, flagged as \texttt{synflag = N}, is likely affected by the same molecular blends as the 508~nm transition and could be reliably fitted in only two stars. Although it is marked with a black circle in several panels of Fig.~\ref{fig:lines_new}, it often suffers from continuum normalization issues and shows significant scatter relative to the other used lines (488.368, 508.742 and 520.041 nm) and across different spectrographs (e.g., \hiprepIII\ and \hiprepVI), owing to its weakness and sensitivity to continuum placement. Examples are shown and discussed in Appendix~\ref{appendix:lines}.

For more than half of the RepGBS sample, the final A(\ion{Y}{II}) abundances are derived from at least four lines (see Table~\ref{tab:lines}). These lines are often used in the literature \citep[e.g.,][]{2023Storm,2023Alexeeva}. Our selected measurements are shown in Fig.~\ref{fig:yII}. We note that different transitions and instruments have excellent agreement, resulting in abundances with high internal precision. The only, though not unexpected, exception is the metal-poor \hiprepIV, for which the weak Y lines result in larger uncertainties and some scatter. Nevertheless, the measurements are broadly consistent, although higher S/N would still be needed for a more robust determination. The only star for which no Y lines were selected is the \typerepV\ \hiprepV; only the reddest line at 554 nm was more or less identifiable, but \texttt{iSpec} failed to reproduce its broadening. 

\subsubsection{Zirconium}

As already evident from Fig.~\ref{fig:lines_new}, zirconium (Zr) is a particularly challenging element to measure, as both \ion{Zr}{I} and \ion{Zr}{II} show large spread in the measurements. A significant number of lines are either weak (blue symbols) or exhibit a large line-to-line scatter (cross symbols). This indicates that Zr lines are often weak or blended \citep{2011Caffau,Heiter21}.

In our sample, we observe a clear dependence of Zr line strength on both metallicity and temperature: Zr lines weaken significantly in hotter stars and become increasingly difficult to detect in more metal-poor candidates. This trend likely reflects the combined effects of the ionization balance and excitation potential. In hotter atmospheres, neutral Zr is largely ionized, causing \ion{Zr}{I} lines to disappear, while in metal-poor stars, the lower electron density and overall opacity weaken both neutral and ionized lines, reducing their detectability \citep{2014Siqueira,2022Kolomiecas}. 

\begin{figure}[t]
    \centering   \includegraphics[width=0.99\linewidth]{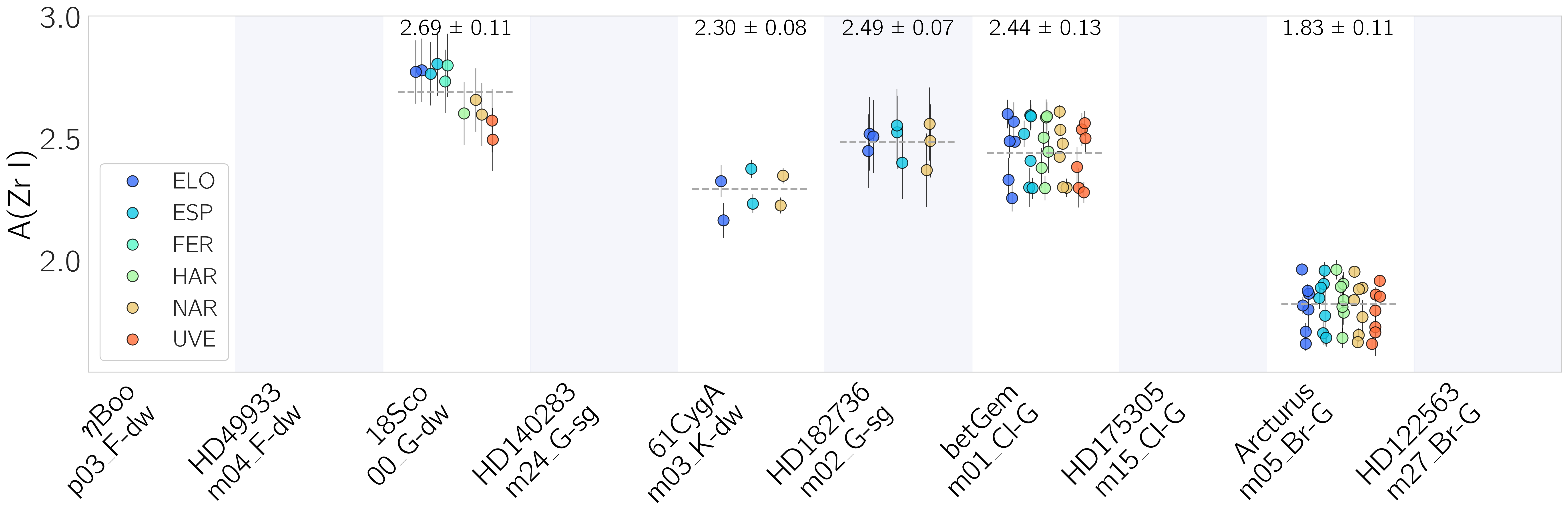}
\includegraphics[width=0.99\linewidth]{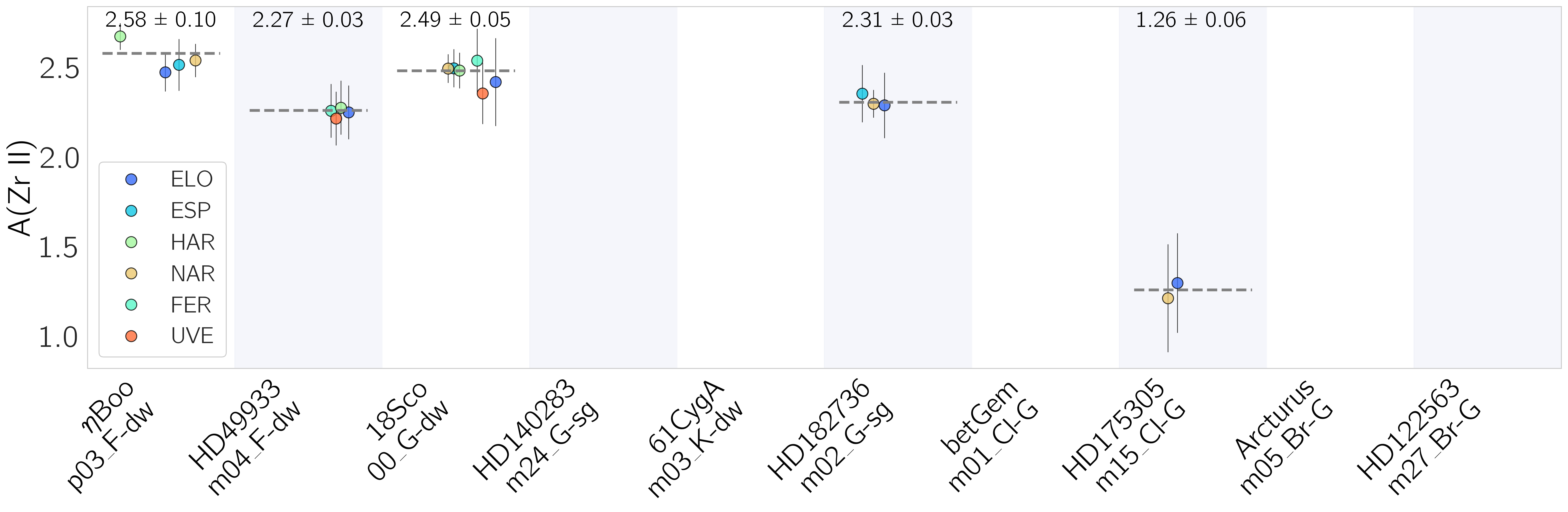}
    \caption{Same as Fig.~\ref{fig:yII} but for \ion{Zr}{I} (upper panel) and \ion{Zr}{II} (lower panel).}  \label{fig:Zr}
\end{figure}

This behavior is particularly evident for the two metal-poor RepGBS (\hiprepII\ and \hiprepVIII), one of which (\hiprepII) is also among the hottest stars in the sample. In these cases, none of the \ion{Zr}{I} lines yield reliable fits, and Zr abundances can be derived only from \ion{Zr}{II}. A similar situation occurs for the hottest dwarfs, \hiprepI\ and \hiprepII, where all \ion{Zr}{I} transitions are too weak or uncertain to be measured. For these stars, only the Zr2\_511.227 line could be used, albeit with noticeable dispersion among spectra (see the upper panel of Fig.~\ref{fig:Zr}). This line is weak and highly sensitive to continuum placement, likely because of a nearby \ion{Fe}{I} feature on its red wing. The varying strength of this blend across the RepGBS can affect the local normalization and introduce systematic uncertainties in the measured abundances \citep{Heiter21}. At the opposite end of the parameter space, the coolest star in the sample, \hiprepV, also poses difficulties. Despite several Zr lines being flagged as usable in Fig.~\ref{fig:lines_new}, the line-by-line dispersion reaches $\sim0.08$ dex (Fig.~\ref{fig:Zr}). For this star, we ultimately rely on the two \ion{Zr}{I} lines at 612 and 613~nm.

The challenges of determining Zr abundance are further illustrated by the \typerepV\ (\hiprepV), \typerepVII\ (\hiprepVII)\ and \typerepIX\ (\hiprepIX), which exhibit the most complete sets of measurable Zr lines. Although Fig.~\ref{fig:lines_new} suggests good consistency among individual transitions, Fig.~\ref{fig:Zr} reveals an intrinsic line-to-line scatter up to 0.13 dex, even among measurements from the same instrument (same colors). This dispersion likely reflects residual systematics related to line weakness, blending, and continuum placement. A similar behavior (even if with smaller dispersion) is observed for \hiprepIII\ in the \ion{Zr}{II} measurements. Finally, the \ion{Zr}{I} 507.825~nm line was discarded for all stars due to its weak and asymmetric profile, consistent with the absence of quality flags in the GES line list.  

The two panels of Fig.~\ref{fig:Zr} show differences of up to 0.18~dex between A(\ion{Zr}{I}) and A(\ion{Zr}{II}) for \hiprepIII\ and \hiprepVI, which are the only two stars for which we could determine Zr abundances for both ionization stages. We note these stars exhibit large uncertainties due to the dispersion among individual line-by-line measurements. Overall, the differences between the two ionization stages in these two stars are comparable to those reported in the literature for galactic abundances \citep{2010Velichko,2017Mena}.

Our analysis shows that to determine Zr abundances the choice of suitable lines strongly depends on both spectral type and metallicity, requiring careful line selection on a star-by-star basis.  Despite these difficulties, we are able to provide Zr abundances for eight stars, leaving out the two most metal-poor ones.

\subsubsection{Molybdenum}
\begin{figure}[t]
    \centering   \includegraphics[width=0.99\linewidth]{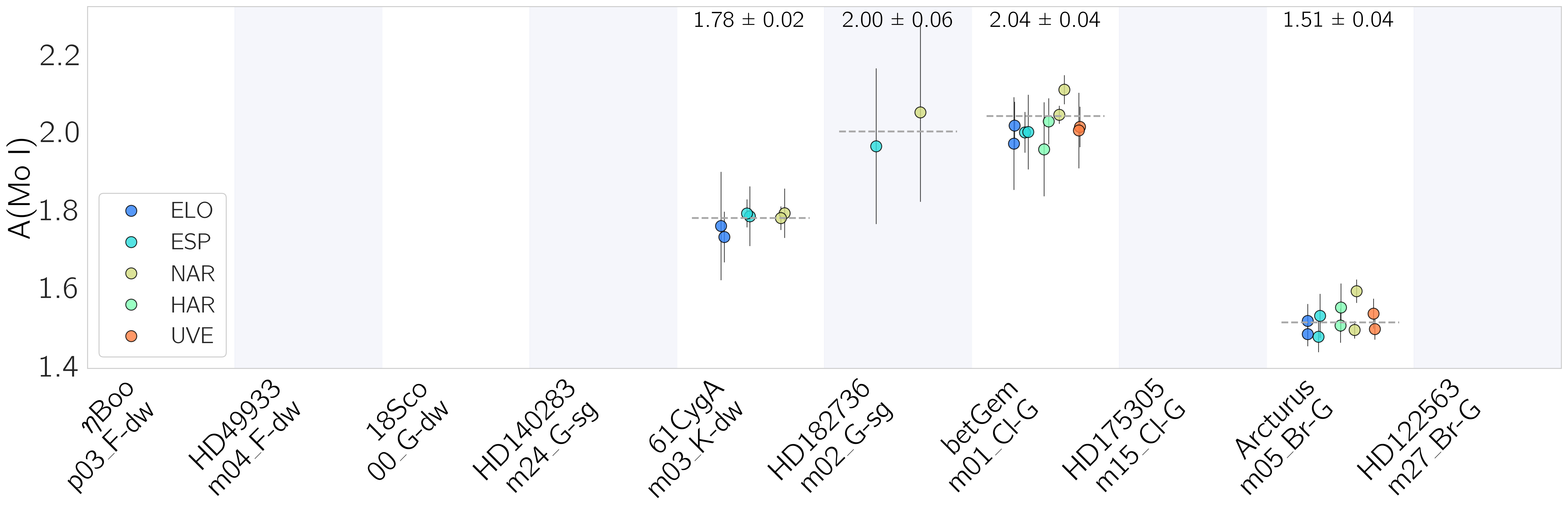}
    \caption{Same as Fig.~\ref{fig:yII} but for \ion{Mo}{I} abundances.}  \label{fig:Mo}
\end{figure}

Molybdenum is not a pure n-capture element, as a significant fraction of its solar abundance ($\sim$23\%) is attributed to p-process nucleosynthesis, while the remaining material originates from the s-process ($\sim$50\%) and r-process ($\sim$27\%) \citep{2020MNRAS.491.1832P}. We analyzed three \ion{Mo}{I} lines commonly used in optical spectroscopic studies \citep[e.g.,][]{2020Mishenina,2022Forsberg,2025Mishenina}. We also attempted to include the 550.6~nm line adopted by \citet{2025Mishenina}, who used VALD in their analysis, but ultimately retained the three original lines, as we found poor synthetic fits for this feature. As noted by \citet{Heiter21}, the \ion{Mo}{I} lines at 557.044 and 603.064~nm are suitable mainly for giant stars, as they are too weak in dwarfs and their hyperfine and isotopic splitting has not been considered.

For the dwarfs, we attempted to estimate the final A(\ion{Mo}{I}) abundances from the line at 553.3~nm, but due to a strong blend with \ion{Fe}{I} we were unable to provide reliable final measurements. The only exception is the \classrepV~\hiprepV, for which the fit appears satisfactory despite the presence of a strong \ion{Fe}{I} blend on the red side. For this star, the Mo1\_557.044 line, though marked with a black circle in Fig.~\ref{fig:lines_new}, was removed due to a \ion{Cr}{I} blend affecting its right wing. 

In the giants the bluer \ion{Mo}{I} line at 553.3~nm is usually too shallow compared to other transitions and can therefore be neglected; despite being marked as good (black circles) in some giants such as \hiprepVII\ and \hiprepIX, it yields lower abundances and unreliable fits. Generally, for giants we can reliably use the Mo1\_557.044 and Mo1\_603.064 lines, whereas for the metal-poor \hiprepVIII\ and \hiprepX, all transitions are too weak to be considered. 

Finally, the difficulty of measuring this element across all ten RepGBSs is further illustrated in Fig.~\ref{fig:Mo}, which highlights the lack of measurements, yet provides two new measurements for \hiprepV\ and \hiprepVI. In particular, the panel for \hiprepVI\ shows only two measurements, as the ELODIE spectra had to be excluded due to their large uncertainties.

\subsubsection{Barium}

\begin{figure}[t]
    \centering   \includegraphics[width=0.99\linewidth]{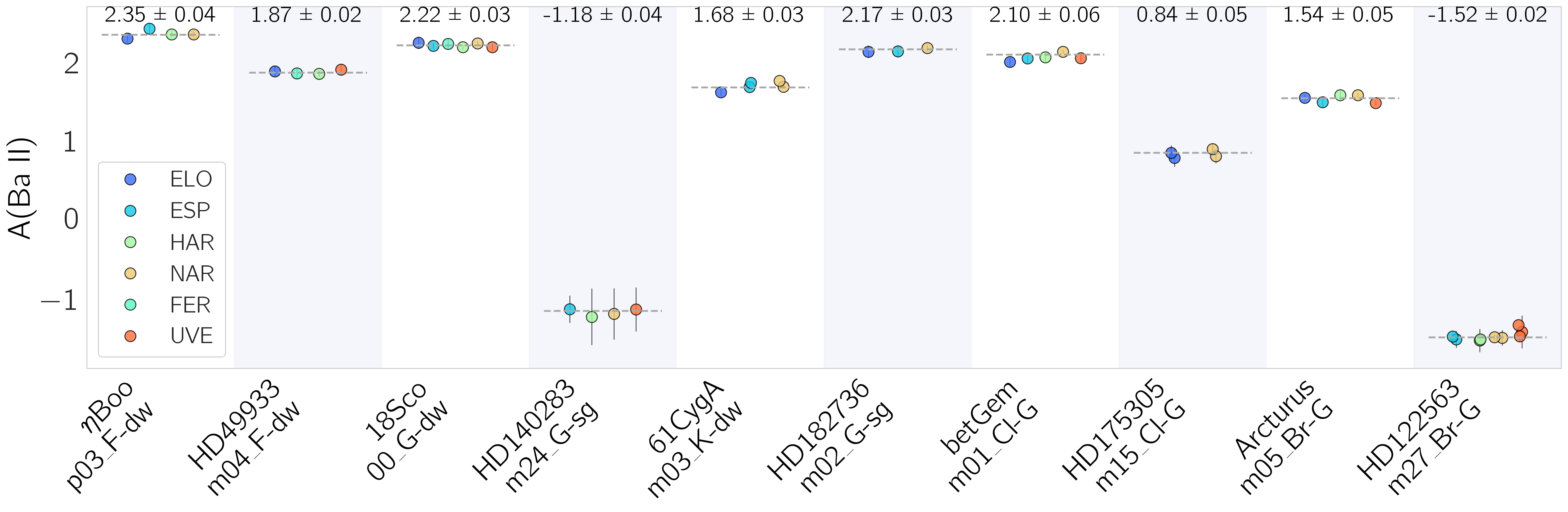}
    \caption{Same as Fig.~\ref{fig:yII} but for \ion{Ba}{II} abundances.}  \label{fig:Ba}
\end{figure}

Because barium is a key benchmarking element in Galactic and stellar evolution studies \citep[e.g.,][]{2023Escorza,2024Guiglion,2025Chen} and a strong tracer of slow ($s$)-process enrichment \citep{1999Busso,2011Bisterzo,2014Karakas}, we carefully derive \ion{Ba}{II} abundances for all RepGBS, despite the fact that its lines can be very strong and are known to be affected by saturation effects \citep[e.g.,][]{Korotin15, 2025Chen}.

We employed three \ion{Ba}{II} transitions which, although widely used in optical abundance studies, are quite strong (except in the most metal-poor stars; \citealt{Heiter21}) and affected by saturation and hyperfine splitting (HFS) \citep{2003Mashonkina,2020Liu,2024Yang}. Among them, \citet{2019Eitner} showed that the Ba2\_585.367 line is the least sensitive to non-LTE effects. We therefore retained this transition for all RepGBS except the metal-poor representative, \classrepIV\ \hiprepIV, which behaves differently from the rest of the sample. In this star, the 585.4~nm and 649.7~nm lines are too weak, while the 614.2~nm line, usually discarded in other stars because of saturation, provides a reliable fit and reasonable EW values. For the same reasons, it can also be adopted for the other metal-poor RepGBS, namely \hiprepVIII\ and \hiprepX.

Not unexpectedly, in giant stars the lines are stronger and become deeper. Hence, we adopted a more cautious approach in applying the same REW cut used for the other elements. In particular, although the Ba2\_614.171 line is flagged in black for most stars in Fig.~\ref{fig:lines_new} and therefore lies within the adopted REW limit, we exclude it from the analysis when it becomes very strong and its core is not well fitted, as illustrated by some examples in Fig.~\ref{fig:yba_panels}. This occurs in the giant and more metal-rich RepGBS. This transition is also known to be blended with a nearby \ion{Fe}{I} line and is among the most problematic of the three commonly used \ion{Ba}{II} lines \citep{Korotin15, Gallagher20}. A clear example is provided by the RepGBS of the \classrepIX\ group. For this star, \hiprepIX, all three Ba lines are flagged as good and fall within the REW range, although with values very close to the saturation limit (REW $\approx -4.55$). However, only the first line is retained (REW $\approx -4.7$), as the others show signs of saturation, especially in the line core.

The reddest \ion{Ba}{II} line at 649.689~nm, is used only when it is not affected by telluric absorption. In fact, the automatic masking of telluric contamination in {\tt iSpec}, used for our spectral library \citep[see][for details]{2025Casamiquela}, has removed this region in \hiprepVI. We retain this line for stars of type \typerepV\ and for the \typerepX\ \hiprepX, where it does not violate the REW criterion owing to the intrinsic weakness of Ba lines at low metallicity, and its line profile remains reliable. In these cases, it shows good agreement with the A(\ion{Ba}{II}) abundances derived from the other line at 585.367 nm. For the other giant stars, however, it exhibits an offset from the median of the line-by-line measurements, likely due to stronger non-LTE effects in giants, which are expected to be less severe in main-sequence stars \citep{2019Eitner}. The final abundances of \ion{Ba}{II} are illustrated in Fig.~\ref{fig:Ba}. It demonstrates that we are able to derive Ba abundances for all RepGBS with a good precision and internal consistency. 
  
\subsubsection{Lanthanum}
\begin{figure}[t]
    \centering   \includegraphics[width=0.99\linewidth]{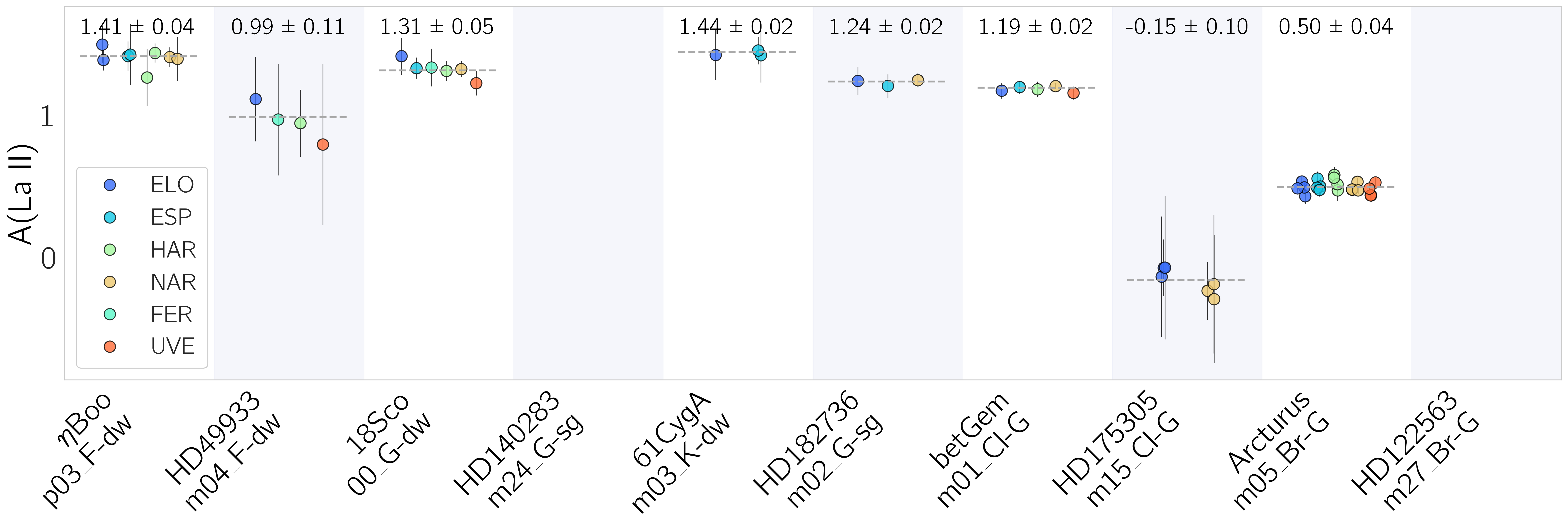}
    \caption{Same as Fig.~\ref{fig:yII} but for La II abundances.}  \label{fig:La}
\end{figure}

Lanthanum abundances in this work rely on five \ion{La}{II} absorption lines, several of which are commonly used in the literature \citep[e.g.,][]{2018MKarinkuzhi,2024Vitali,2025Lombardo}. We include these transitions even when GES flags are not available, based on their empirical performance in the RepGBS sample. Among them, the 511.456 nm line is the most frequently measurable across the sample, although it remains undetectable in dwarfs and is generally weak in most spectra. The lines La2\_480.403 and La2\_639.046 have positive $\log gf$ flags, while the remaining transitions lack flag information. Moreover, these two lines include three and four HFS components in the GES line list, respectively. It should be noted that the GES line list includes HFS only for flagged lines. For transitions without GES flags, HFS information is not provided. In cases of blends, such as for the cool \typerepV\ star or the \typerepIII\ \hiprepIII, \texttt{iSpec} correctly fits both the line and the blend, and discards the affected regions by applying appropriate line masks around the line profile. For \hiprepV, as shown in the fifth panel of Fig.~\ref{fig:La}, fewer spectra are available compared to the other elements because the NARVAL observations had to be discarded due to quality issues in this spectral region.

Using a combination of these transitions, we derived final La abundances for 80\% of the RepGBSs. This excludes the two metal-poor cases (\hiprepIV\ and \hiprepX), where the lines are too weak. In addition, for the \typerepII\ star \hiprepII, the intrinsic weakness of the lines is reflected in the large uncertainties shown in Fig.~\ref{fig:La}.

\subsubsection{Cerium}
\begin{figure}[t]
    \centering   \includegraphics[width=0.99\linewidth]{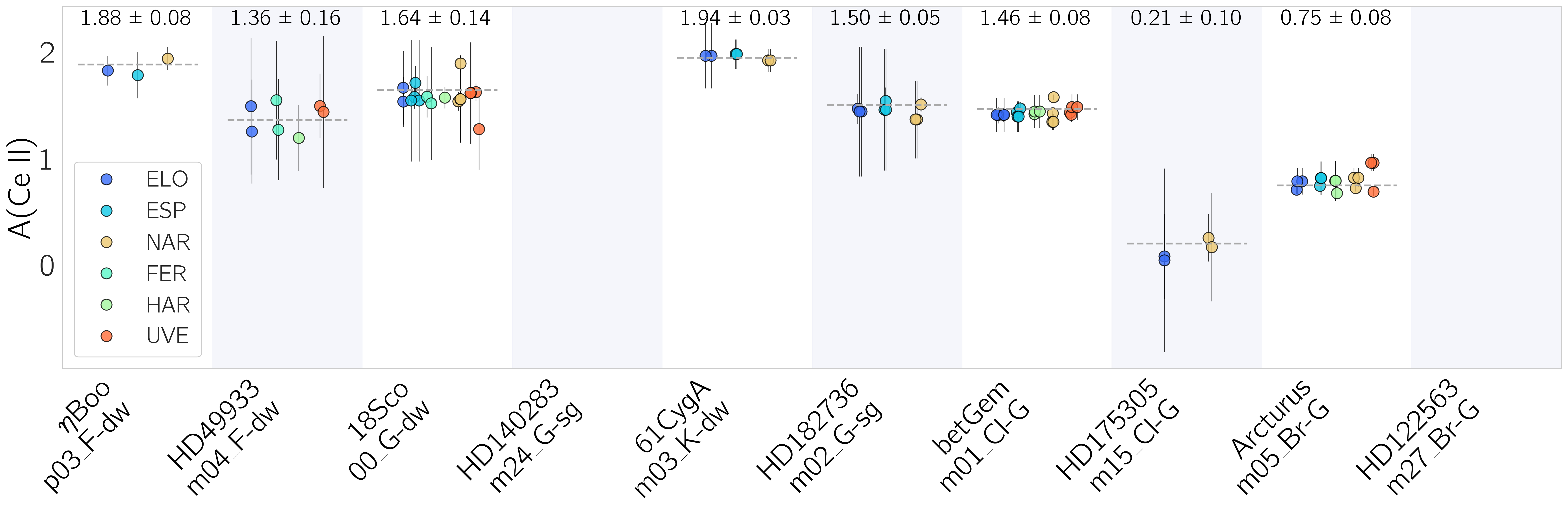}
    \caption{Same as Fig.~\ref{fig:yII} but for \ion{Ce}{II} abundances.}  \label{fig:Ce}
\end{figure}

We began by examining the \ion{Ce}{II} lines at 527.4, 533.1, and 604.3~nm, which have reliable $\log gf$ values according to \cite{Heiter21}. Following their recommendations, we avoided the bluer \ion{Ce}{II} lines at 498.4 and 504.4~nm due to strong blends, and instead adopted the three transitions listed above, which are also commonly used in the literature \citep{2018MKarinkuzhi,2023Contursi}. 

A particular case is the Ce2\_561.025 line, which does not have GES flags, but was included based on previous studies \citep[e.g.,][]{2007Gratton,2009Lawler,2013Mishenina}. It is employed for only two RepGBS, but for one of them, namely the \typerepV, it is the sole usable feature to derive the final A(Ce), as the other Ce lines are either blended or affected by continuum suppression, producing scattered and unreliable line-by-line abundances. Furthermore, it is adopted for \hiprepVII, yielding results consistent with those obtained from the other transitions. Although this line is also visible in other RepGBS, it often leads to overestimated abundances and is therefore not used in the final analysis. For the \typerepII\ and \typerepX\ metal-poor stars, the lines are too weak to provide reliable fits. In contrast, for \classrepVIII\ \hiprepVIII\ the selected transitions, although weak, can still be used to derive consistent abundances, with the increased uncertainties reflected in Fig.~\ref{fig:Ce}. For some metal-rich stars (e.g., \hiprepI\ and \hiprepIII), the Ce2\_527.423 line is strong but affected by continuum-normalization issues on the red wing and a nearby \ion{Cr}{I} blend, which may influence the measurement and must be properly accounted for in the synthesis. Nevertheless, the abundances derived from different spectrographs remain mutually consistent within the uncertainties. A similar continuum-normalization issue, compounded by a nearby \ion{Cr}{I} blend, is observed for \hiprepVII.

The Ce2\_604.337 line is not usable for the majority of the RepGBS sample, as it is generally weak and thus yields large uncertainties. We therefore retained it only for stars in which it provides results consistent with other lines and for which the fit remains reliable despite its weakness. In practice, it is used in three cases, including \hiprepIII, \hiprepVI\ and \hiprepIX. The effect of line weakness is naturally reflected in the uncertainties computed by \texttt{iSpec} (see the corresponding panels in Fig.~\ref{fig:Ce}). For \hiprepIX, we rely on the bluer and redder transitions, as the central Ce2\_533.056 line shows normalization issues.

\subsubsection{Praseodymium}
\begin{figure}[t]
    \centering   \includegraphics[width=0.99\linewidth]{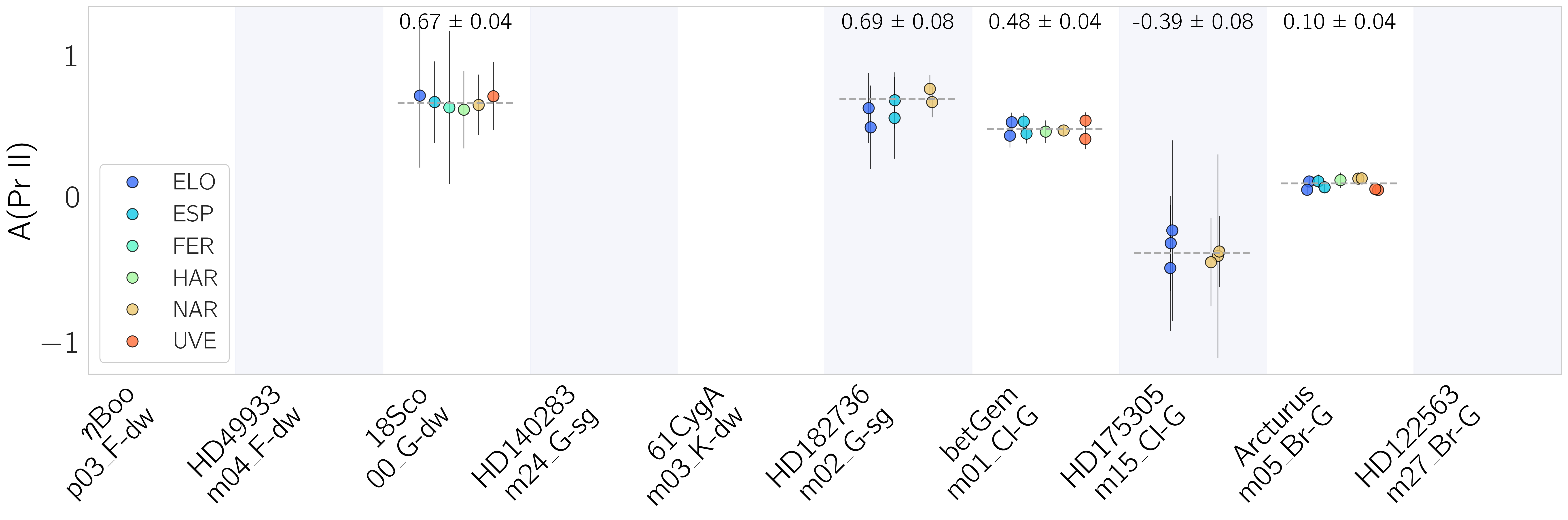}
    \caption{Same as Fig.~\ref{fig:yII} but for \ion{Pr}{II} abundances.}  \label{fig:Pr}
\end{figure}

\ion{Pr}{II} abundances were derived using three commonly adopted optical lines for heavy element analyses \citep[e.g.,][]{2009ASneden,2018MKarinkuzhi}. For \hiprepI\ and \hiprepII, although two transitions are flagged as usable, albeit weak (crosses) in Fig.~\ref{fig:lines_new}, the lines are in practice either extremely weak or entirely absent, preventing reliable abundance determinations. For the \classrepIII\ group (\hiprepIII), only one of the \ion{Pr}{II} lines can be retained. Although this reduces the number of available measurements, this choice is justified by the large line-to-line scatter shown in the Pr column of Fig.~\ref{fig:lines_new} for this RepGBS. For the metal-poor stars, all \ion{Pr}{II} features are generally too weak to provide meaningful constraints. Similarly, for the \typerepV\ \hiprepV\ and the \typerepX\,\hiprepX, no reliable \ion{Pr}{II} abundances could be obtained, despite one line being formally tagged as usable. Moreover, we note that the Pr2\_513.514 line, which is not flagged in the GES line list and therefore has no HFS information included (as for other unflagged lines, similarly to the case discussed for La), systematically yields abundances that deviate by up to $\sim 0.5$ dex from the mean of the other line measurements. In giant stars, we attribute this discrepancy to uncertainties in continuum placement caused by blending with nearby \ion{Fe}{I} features. This issue is observed across all giants in the sample, with the exception of \typerepVIII\ (\hiprepVIII), for which all three \ion{Pr}{II} lines exhibit reliable continuum normalization and, despite being weak, can be successfully fitted and yield consistent results. 

In summary, due to the shallow line depths and the associated measurement difficulties, we are able to derive Pr abundances for only about 50\% of the RepGBSs. These are primarily the giants and the more metal-rich objects, where the lines are sufficiently strong.

\subsubsection{Neodymium}
\begin{figure}[t]
    \centering   \includegraphics[width=0.99\linewidth]{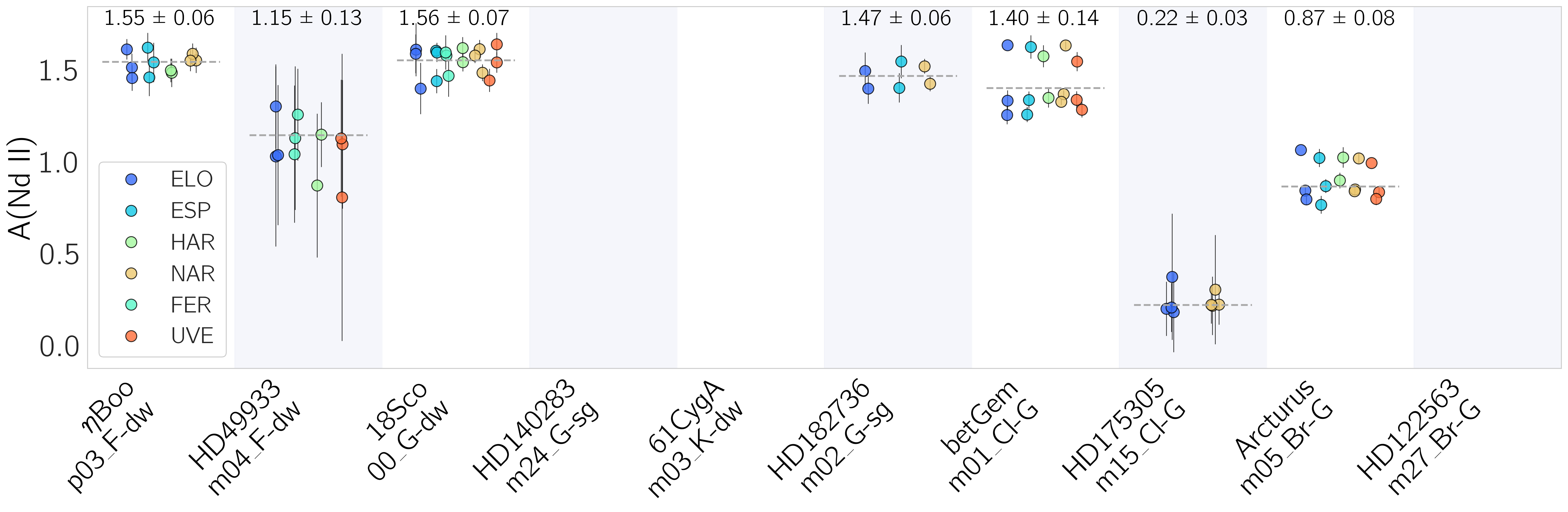}
    \caption{Same as Fig.~\ref{fig:yII} but for \ion{Nd}{II} abundances.}  \label{fig:Nd}
\end{figure}

Neodymium, a tracer of both $s$- and rapid($r$)-process nucleosynthesis \citep{1999Arlandini}, has been extensively investigated in both the optical and near-infrared regimes \citep[e.g.,][]{2016Hasselquist,2025Racca,2025Dixon}. In the selection of the Gaia-ESO (GES) line-list presented by \citet{Heiter21}, \ion{Nd}{II} transitions were carefully evaluated, including the effects of HFS and isotopic ratios, in addition to the standard flags for blending and uncertainties in $\log gf$.

In this work, we started considering six \ion{Nd}{II} lines. Among these, the Nd2\_518.117 and Nd2\_521.565 transitions lack both \texttt{synflag} and \texttt{gfflag} estimates and were never selected for any of the RepGBSs (see Table~\ref{tab:lines}). The redder Nd2\_568.852 line also does not have flags reported in the GES line list and hence no HFS information, but it was nevertheless measurable and retained for three RepGBSs.

Even though literature studies often rely on bluer transitions, as is typical for n-capture element abundance analyses \citep[e.g.,][]{2018Sakari,2024Bandyopadhyay,2025Griffith}, in this work, given the wavelength range investigated, the \ion{Nd}{II} lines at 513.059, 529.316, and 531.981 nm are the most frequently used. However, the 513.059 nm line cannot be employed for all stars because it is blended on its blue wing with an \ion{Fe}{I} feature (indeed flagged with \texttt{synflag = N}), making its reliability strongly dependent on the quality of the spectral synthesis and continuum normalization. For instance, for \hiprepIX\ we adopted the three reddest transitions. The Nd2\_513.059 line, although formally flagged as usable, suffers from normalization issues in a very crowded spectral region and was therefore excluded. It is also worth noting that, despite being flagged by GES, no HFS data are available for this transition in the GES line list, and therefore HFS is not included.

\hiprepV\ represents another case that required careful evaluation. In fact, although two lines are flagged as good in Fig.~\ref{fig:lines_new}, no reliable synthetic fit and no final A(\ion{Nd}{II}) abundance could be obtained. Similarly, for \hiprepVI, although more lines than the two selected are marked as good, they cannot be selected. In particular, Nd2\_529.316 is affected by a blend on its blue wing and was therefore discarded, while Nd2\_568.852 is contaminated by telluric absorption. The line at 518.117~nm is discrepant due to its shallowness and blending, showing large deviations from $\langle A(x) \rangle$ in many RepGBS, leading to imprecise fits. It was therefore excluded along with the similarly unflagged transition Nd2\_521.565. The \hiprepVIII\ representative exemplifies the weakness and unreliability of these lines. Although several \ion{Nd}{II} lines flagged as good were retained to maximize the available measurements for this star, the 518.1 and 521.6~nm transitions were excluded due to their weak and shallow profiles.

Overall, Nd abundances are obtained for about 70\% of the RepGBS sample, usually from several lines, although with a larger line-to-line scatter than for other elements. The scatter is particularly noticeable for some stars, such as the hottest object in the sample, \hiprepII\ ($\sigma(\mathrm{Nd}) \approx 0.13$ dex), and the cooler \hiprepVII\ ($\sigma(\mathrm{Nd}) \approx 0.14$ dex).

\subsubsection{Europium}
In the wavelength range explored by the GES line list, many \ion{Eu}{II} transitions are weak or partially blended \citep{Heiter21}. The Eu2\_664.510 line investigated here has \texttt{gfflag = Y}. However, it is affected by HFS, which must be properly accounted for in the spectral synthesis, as is done in the present analysis. In some cases, the \ion{Eu}{II} line at 643.8 nm has also been employed \citep[e.g.,][]{2025Santarelli}, but it is not always detectable and is heavily blended with a nearby silicon transition \citep{2000Mashonkina,Heiter21}. Consequently, many studies focusing on $r$-process elements combine this feature with additional \ion{Eu}{II} lines, most of which are located in the bluer part of the spectrum \citep[e.g.,][]{2022Hansen,2025Guo,2025Racca}.
\begin{figure}[t]
    \centering   \includegraphics[width=0.99\linewidth]{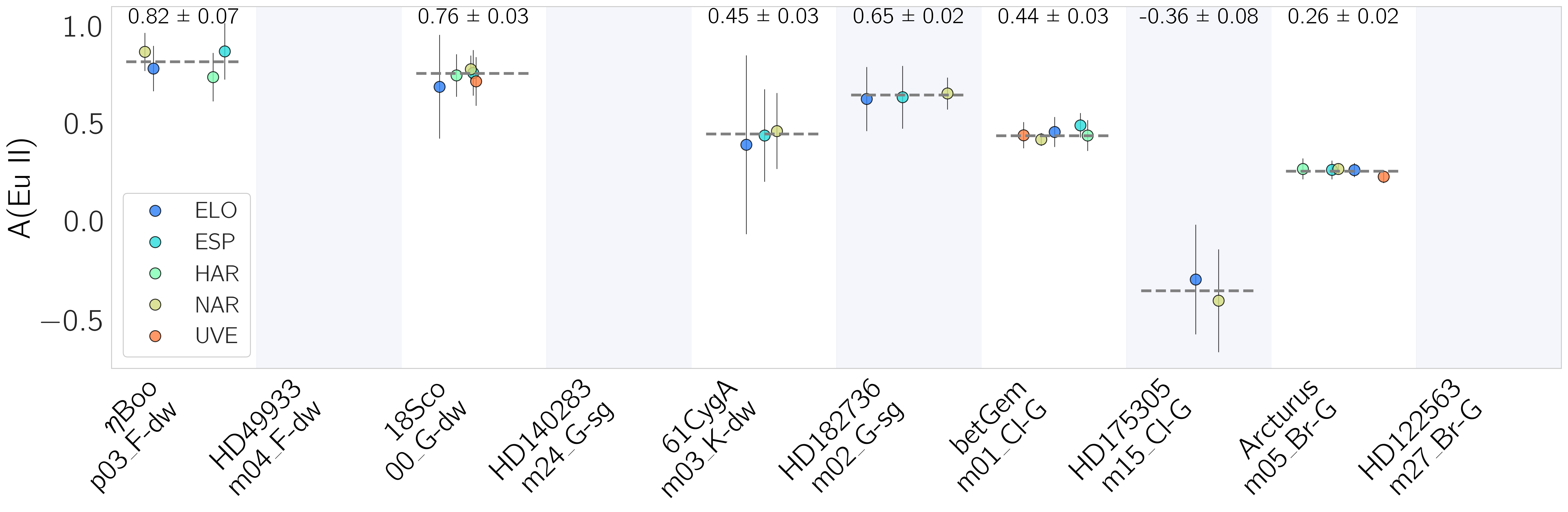}
    \caption{Same as Fig.~\ref{fig:yII} but for \ion{Eu}{II} abundances.}  \label{fig:Eu}
\end{figure}

In the present analysis, relying on a single, intrinsically weak absorption line limits the determination of the final Eu abundance, A(\ion{Eu}{II}), to seven of the RepGBS (see Fig.~\ref{fig:Eu}). In particular, Eu abundances could not be derived for stars in which this line was too weak to be reliably fitted. This is the case for the metal-poor representatives \hiprepII, \hiprepIV, and \hiprepX. Indeed, for metal-poor stars, Eu is more commonly derived using transitions in the 380–420~nm wavelength range, where stronger and less blended lines are available \citep{2002Sneden,2003Aoki,2025Racca}.

For the remaining cases, such as \hiprepI\ and \hiprepV, the Eu2\_664.520 line, although shallow, can still be fitted in a reliable manner. Nevertheless, as illustrated in Fig.~\ref{fig:yba_panels} for some stars this transition is blended with a nearby \ion{Fe}{I} line. This blending must therefore be carefully accounted for in the spectral synthesis and abundance determination.

\subsection{Comparison with literature}\label{sect:literature_comp}

\begin{figure}[t]
    \centering   \includegraphics[width=0.99\linewidth]{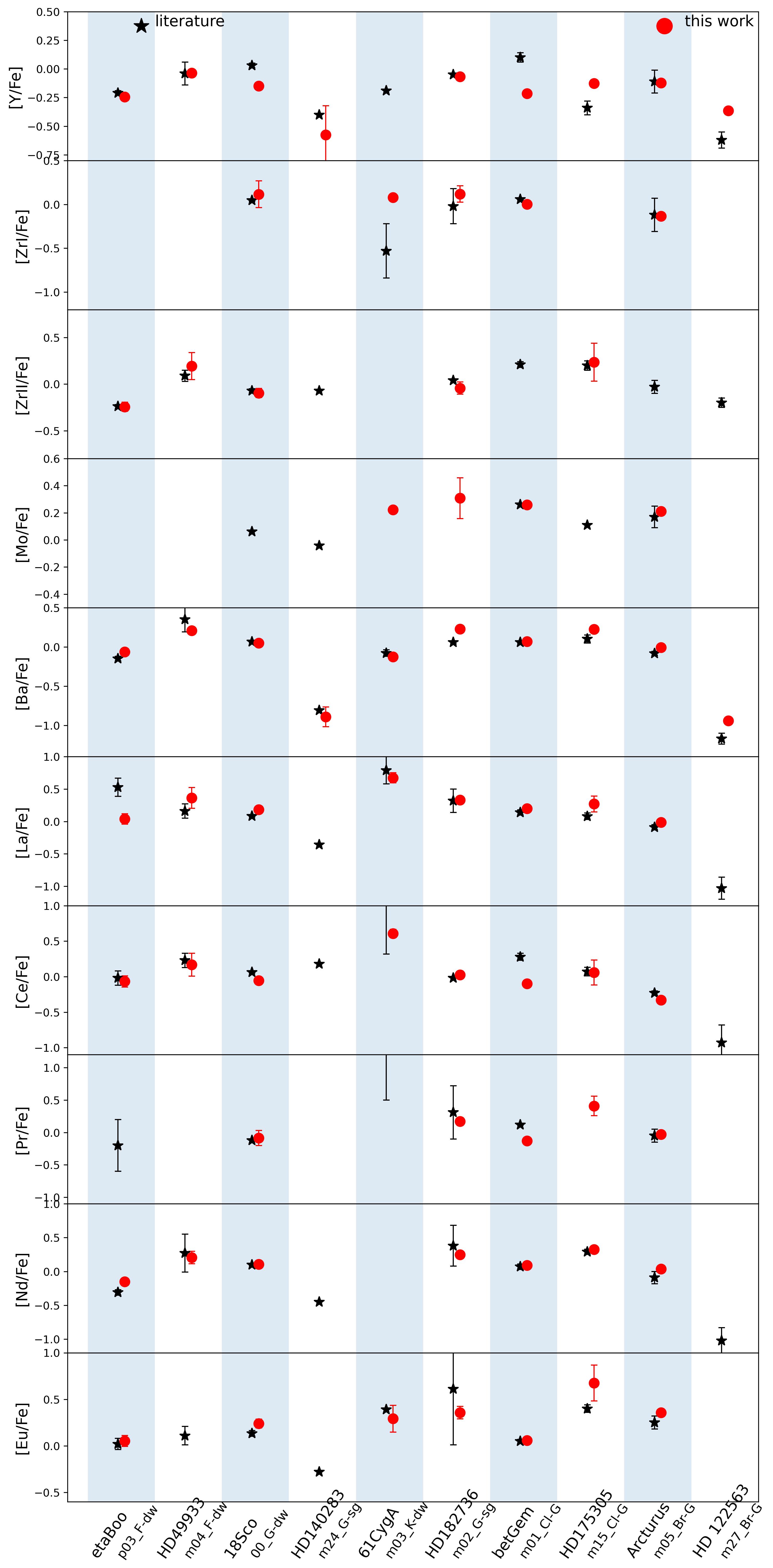}\caption{Literature comparison for the RepGBS abundances. Black stars correspond to literature values, while red circles represent the abundances derived in this work. In some panels for the star \hiprepV\ (Ce and Pr), the literature measurements are not shown because they lie far above our abundance range. We exclude them from the figure for better visibility.}  \label{fig:litgbs}
\end{figure}

Figure~\ref{fig:litgbs} shows a comparison between our results, derived as described above, and literature values. Our abundance ratios are expressed relative to Fe, adopting the solar abundances from \cite{grevesse07}, while for the literature values we adopted the solar abundances reported in the corresponding references. The compiled literature abundances and their corresponding references are listed in App.~\ref{app:lit}, alongside a brief description of the main assumptions and methods of each study. Because the literature compilation combines different line selections, analysis techniques, solar abundance scales, and wavelength coverages, some level of discrepancy is expected. In particular, several literature abundances are derived from transitions outside the spectral range analyzed in this work. 

Overall, we find good agreement with the literature. The comparison also highlights the intrinsic difficulty of measuring n-capture elements, as abundances are not available for all elements in all stars. This is particularly evident for cool, active or metal-poor stars, where line blending, limited wavelength coverage, and stellar properties such as rotation further complicate the analysis. An example is the \typerepV\ \hiprepV, for which only five elements overlap with our analysis. This is unsurprising given that it is a cool, variable, and magnetically active star whose activity cycle can significantly affect spectral line profiles \citep{2018Boro}.

For the metal-poor stars, we were able to measure only a limited number of heavy elements, and the literature coverage is likewise incomplete, despite these objects being extensively studied. One of the metal-poor targets, namely \hiprepII, is also a fast rotator, with a projected equatorial rotational velocity $v \sin i$ estimated in the literature to range between 11 and 18 $\mathrm{km\,s^{-1}}$ \citep{2023A&A...676A.129H}, which may introduce additional difficulties in the abundance determination. The other fast rotator in the sample is \hiprepI, with a $v \sin i$ of $11-13\,\mathrm{km\, s^{-1}}$ \citep{2019Buder,2023A&A...676A.129H}. For this star, we find discrepancies of $\sim 0.1$~dex for \ion{Ba}{ii} and \ion{Nd}{ii}, and of $\sim 0.5$~dex for \ion{La}{ii}, while the abundances for four other elements agree very well. This further supports that \texttt{iSpec} adequately accounts for rotational broadening through the line-broadening parameters adopted in the spectral fitting. 

The agreement for Ba, Pr, Nd, and Eu is remarkable, considering the broad range of spectral types considered and the diversity of literature studies included in the comparison. Ba is the element with the most complete comparison. 

The metal-poor stars, \hiprepII\ and \hiprepX, show differences of $\sim 0.17$ dex. For the former, the discrepancy remains within the error bars, while for \hiprepX\ the offset lies outside the uncertainties and may be due to differences in the adopted atmospheric parameters and the use of the EW method, as outlined in Appendix~\ref{app:lit}. Similar reasons likely explain the other large difference found for Ba ($0.19$ dex) in \hiprepVI. The overall good agreement may indicate that, despite Ba being an element affected by non-LTE effects \citep{Korotin15,Gallagher20}, consistent LTE/non-LTE assumptions lead to comparable results. Moreover, the strength of its lines, which are usually free of significant blends, makes Ba a robust proxy for measuring n-capture element abundances.

For Pr, although we could not identify suitable lines for the \classrepI\ and \classrepV\ groups, abundances for \hiprepI\ and \hiprepV\ are reported by \cite{2017AJ....153...21L}, who analyzed spectra covering a wavelength range similar to ours. However, the associated uncertainties are large, indicating that these measurements are highly uncertain, which supports our decision to reject these lines for these stars. Indeed, although the online tables of \cite{2017AJ....153...21L} list Pr abundances for these targets, this element is not discussed in the analysis presented in that paper. We did not find Pr measurements in the literature for \hiprepII\ or for the metal-poor stars \hiprepIV\ and \hiprepX; similarly, we were also unable to derive reliable abundances for these stars. 

For Nd, the usual “problematic” RepGBS cases prevent us from providing a final A(Nd) for the metal-poor \hiprepIV\ and \hiprepX, as well as for the \typerepV\ \hiprepV. This likely reflects the limited blue wavelength coverage of our spectra, which is generally more favorable for heavy element abundance determinations in metal-poor stars. While \cite{2023A&A...676A.129H} do not report the specific lines used, their analysis relies on GES setups (UVES and GIRAFFE) with broader blue coverage, whereas
\cite{2014AJ....147..136R} employ bluer transitions ($410.9$ $-482.4$~nm) for their \ion{Nd}{II} abundance determinations.
We have a similar situation for Eu, as we do not provide Eu abundances for the most metal-poor stars. However, literature values are available for these stars, specifically \cite{2015A&A...584A..86S} for \hiprepIV\ and \cite{2014AJ....147..136R} for \hiprepX. These are based on the Eu line at 412.9~nm, which lies outside the wavelength range considered in this work.

Yttrium is interesting because we use several lines to determine the abundances, yet the differences with the literature are significant for most stars. As an element widely used in Galactic archaeology, it deserves further investigation. We found that \cite{Melendez14} employed very different oscillator strengths for \hiprepIII, even for lines in common with our analysis, in addition to transitions outside our spectral range (e.g., at 460.7~nm). Furthermore, their study uses the EW method and a differential line-by-line analysis. For \hiprepVII, although \cite{2025A&A...701A.153S} used the same spectral range and atomic data as in our analysis, some of the lines in their list were excluded from our work. In addition, the adopted stellar parameters differ, particularly the metallicity, which is $\sim 0.2$ dex higher than our value. For \hiprepI, a direct comparison with \cite{2014AJ....148...54H} is not straightforward, as their study compiles results from different literature sources. Nevertheless, we find very good agreement, despite possible differences in analysis methods and line selection. The metal-poor stars \hiprepVIII\ and \hiprepX\ were analyzed by \cite{2014AJ....147..136R}. While there is a substantial overlap in line selection, their analysis extends to bluer wavelengths. The lines in common adopt the same oscillator strengths as in our work, although differences in stellar parameters may contribute to the abundance discrepancies. We also note the recent discussion of microturbolence-related effects in Y abundances for metal-poor stars by \cite{2025arXiv251213590S}. In particular, the difference in $v_{mic}$ between our analysis and the literature for \hiprepVIII\ exceeds 0.5 km\ s$^{-1}$. 

To summarize, although several lines within our spectral range are available and our results are robust (see Fig.~\ref{fig:yII}), Y abundances show a strong dependence on stellar parameters, while uncertainties in the atomic data lead to notable differences between studies. Consequently, the selection of suitable \ion{Y}{II} lines must be performed on a star-by-star basis, accounting for blending and continuum-placement effects \citep{2017Nissen,2024Guiglion}.

Regarding Zr, we compared the abundances derived from the two ionization stages separately, as our results differ between them, which is a behavior also reflected in the literature. This difference is immediately evident in both the literature compilations and in our RepGBS sample, where several stars lack \ion{Zr}{I} abundances. Nevertheless, we find an overall good agreement with the literature for both ionization stages. Interestingly, the difference in the Zr abundance follows the same trend as the difference observed for Y. A considerable discrepancy ($\sim 0.5$ dex) is found for \hiprepV\ in \ion{Zr}{I}. However, this star is among the most challenging targets in the sample due to its low temperature, which is also reflected in the large uncertainties reported in the literature.

Turning to the ionized transition, literature compilations provide measurements for all the metal-poor RepGBS. For \hiprepX, \cite{2014AJ....147..136R} employs transitions in the blue spectral region, around $414-420$~nm. For \hiprepIV, the abundance is reported in the recent work of \cite{2025A&A...701A.153S}; although their analysis covers a similar wavelength range, it relies on different lines that are not included in our selection. The absence of our measurement for \hiprepIX, analyzed in the same work as \hiprepX\ by \cite{2014AJ....147..136R}, can be explained in a similar way.

The overall picture for La is similar to that found for Zr, with generally good agreement with the literature. An exception is \hiprepI, for which the La abundance reported by \cite{2023A&A...676A.129H} differs from our result, despite the atmospheric parameters being similar. This suggests that the discrepancy likely arises from the different lines adopted, which is also the case for the lowest-metallicity RepGBS targets. In fact, \cite{2025A&A...701A.153S} rely on two \ion{La}{II} lines for \hiprepIV\ at wavelengths $\leq 480$~nm, in addition to lines within our spectral range. We attempted to include these transitions but discarded them because the spectral fits were of insufficient quality. Similarly, for \hiprepX, \cite{2014AJ....147..136R} used a line at 408.6~nm. 

Regarding Ce, no abundances could be derived for \typerepIV\ and \typerepX. For \hiprepIV, \cite{2025A&A...701A.153S} use lines within our spectral range; however, the transitions overlapping with our selection (527.4 and 604.3~nm) were found to be too weak for reliable abundance measurements. For \hiprepX, as for other heavy elements, \cite{2014AJ....147..136R} rely on \ion{Ce}{II} lines at $\lambda < 480$~nm, outside our spectral coverage. For the remaining RepGBS, the agreement in Ce is generally good, except for \hiprepV. This discrepancy, although within the literature uncertainties, likely reflects differences in the adopted atmospheric parameters. The large error bars reported by \cite{2017AJ....153...21L} for \ion{Pr}{II} in the same star suggest similar underlying issues.

Finally, concerning Mo, we derived abundances for only four stars and found literature values for five, with two in common. Only for \hiprepIX\ and \hiprepVII\ we can compare our results, which agree very well. For \hiprepV\ and \hiprepVI, we could not find a comparison value in the literature, suggesting we might be the first ones providing accurate Mo abundances for these stars. For \hiprepIII, \hiprepIV\ and \hiprepVIII, the abundances were derived by \cite{Melendez14}, \cite{2020A&A...638A..64P}, and \cite{2008Roederer}, respectively. They use lines that lie in a bluer wavelength range than our spectra. Notably, for \hiprepIV\ UV lines were analyzed.

\section{Neutron-capture abundances of the GBSv3}\label{sect:4}

\subsection{Validation of line selection for clustering}\label{sect:val_clust}

To determine the abundances for all GBSv3 stars, we first separate them into groups according to the clustering procedure described in Sect.~\ref{sect:repgbs}. The line selection obtained for the RepGBS is then applied to all stars in each group. This procedure is illustrated in Appendix~\ref{appendix:line_sel_clustering}, where Fig.~\ref{fig:line_validation_clusters} shows a plot similar to Fig.~\ref{fig:lines_new}.

Overall, the line selection performs well, with only a few cases where selected lines are red, the main exception being Cluster IX. In this group, some lines are saturated for several elements. This example highlights that the RepGBS adopted from PVIII are not always optimal representatives of the entire GBS spectral dataset. For instance, \hiprepIX, intended to represent a bright giant, has a metallicity of only $\approx -0.55$~dex and lies in a parameter region where effects arise for elements such as Ba, for which the lines tend to be very strong in more metal-rich giants that might be assigned to its group \citep[see also discussion in][]{2024A&A...687A.164V}. In this sense, including an additional representative star would have been useful to better sample the range of metallicities and stellar parameters in the GBSv3. 

However, as extensively discussed in PVIII and in Sect.~\ref{sect:repgbs}, the RepGBS are required not only to represent the different classes of GBS but also to provide a sufficient number of spectra to allow robust assessment of line selection and associated uncertainties. Even more metal-rich giants would still exhibit saturated Ba lines, and the same limitation would remain for this element. We therefore retained the line selection for these clusters and applied the standard REW cut as described in Sect.~\ref{sect:line_sel} to remove the red measurements. This further reduced the number of usable lines, especially for Cluster IX, but did not remove any element from any star due to inadequate line selection except Ba for the most metal-rich giant stars, whose lines would be saturated. As noted in Sect.~\ref{sect:abundances}, even selected lines can present issues in individual spectra and may need to be discarded from the line-by-line analysis, for example, due to normalization problems or low S/N. To mitigate this, we imposed a further constraint on the uncertainties that assesses the quality of the synthetic line fits, ensuring that only reliable measurements contribute to the final abundances. All lines evaluated for the RepGBS in fact satisfy this criterion. 

We finally check the uncertainties of all measurements and find that the majority lie below $\sim 0.2-0.3$ dex (see Fig.~\ref{fig:histo_err}).This tells us that we are able to determine abundances with similar precision and accuracy for all GBS and elements. Except for the very few cases that we have discarded from the final abundances, there is no element or stellar group that has a bias (see App.~\ref{app:errors} for further discussions). This selection results in a final set of measurements per element that varies across the sample, with typical coverage ranging from about 60–85\% of the full GBS sample, reaching 86\% for \ion{Y}{II} (the element with the largest number of available lines) and down to 32\% for the least well-covered species, \ion{Zr}{II}, for which only a single transition is considered.

\subsection{Solar elemental abundances}
\begin{figure}[t]
    \centering   \includegraphics[width=0.99\linewidth]{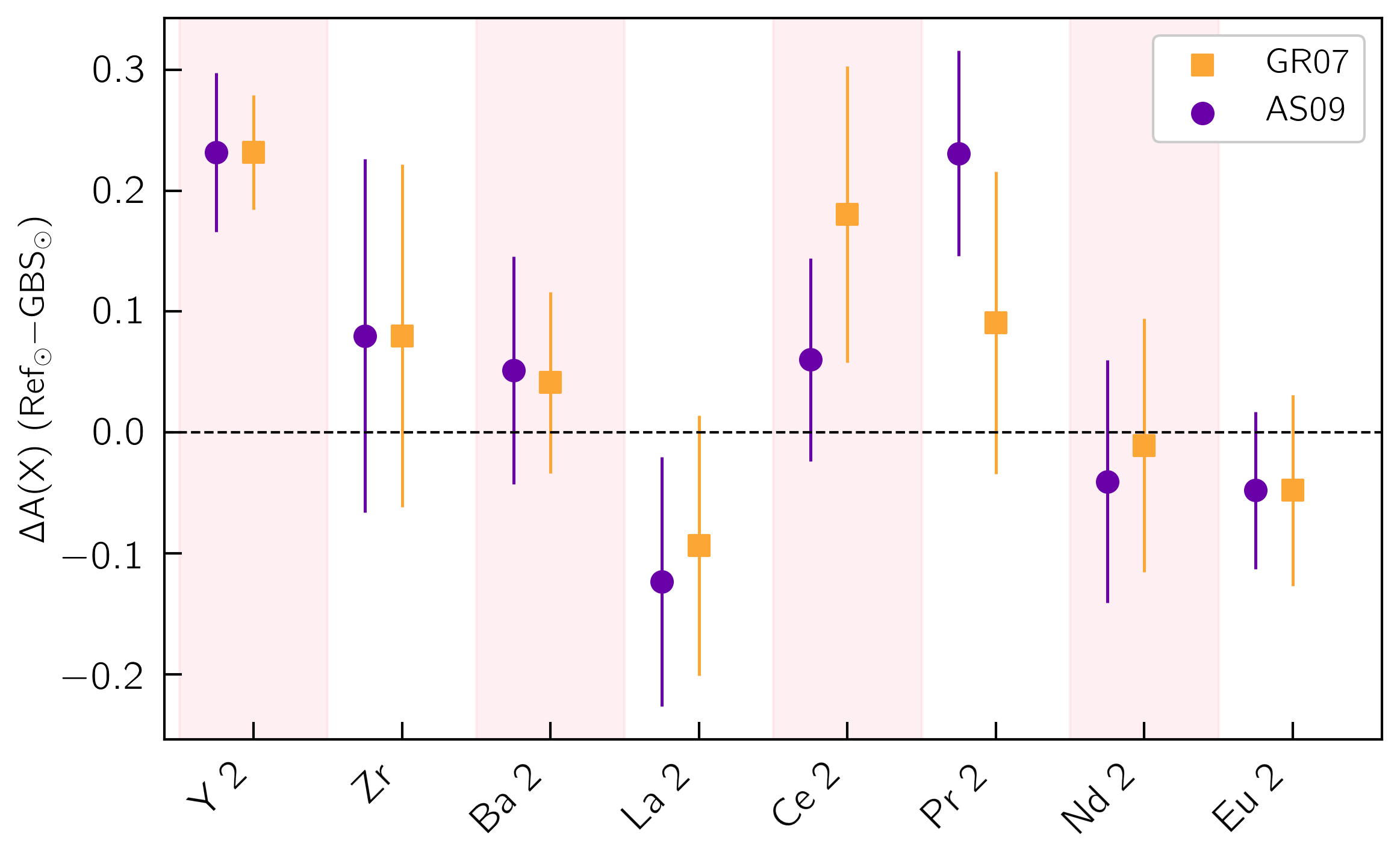}
    \caption{Differences between our solar abundances and the literature values from \cite{grevesse07} (GR07) and \cite{2009Asplund} (AS09). These values are also reported in Table~\ref{Tab:solar}.}
    \label{fig:solar_delta}
\end{figure}

As the most widely studied star, whose physical properties underpin stellar models, nucleosynthesis, magnetic activity, and much more, the Sun has long served as the benchmark for the calibration of stellar atmospheres and abundance analyses \citep{2008Gustafsson,2013Magic,2019Lodders}. It is therefore worthwhile to carry out a separate comparison between our results and those available in the literature. 

For this purpose, we adopt the solar abundance compilation of \citet[][hereafter GR07]{grevesse07}. Although a later work by the same authors \citep{2015AGrevesse} reevaluates the heavy element abundances using 3D hydrodynamical solar model atmospheres, we retain the values from GR07 because most of the literature studies used for comparison in this work rely on this reference scale. For completeness, as also widely employed in the literature, a comparison with the solar abundances of \citet[][hereafter AS09]{2009Asplund}, which presents a 3D-based reanalysis of the n-capture solar abundances based on the work of \citet{2010Grevesse} is also included.
\begin{figure*}[t]
    \centering   \includegraphics[width=0.99\linewidth]{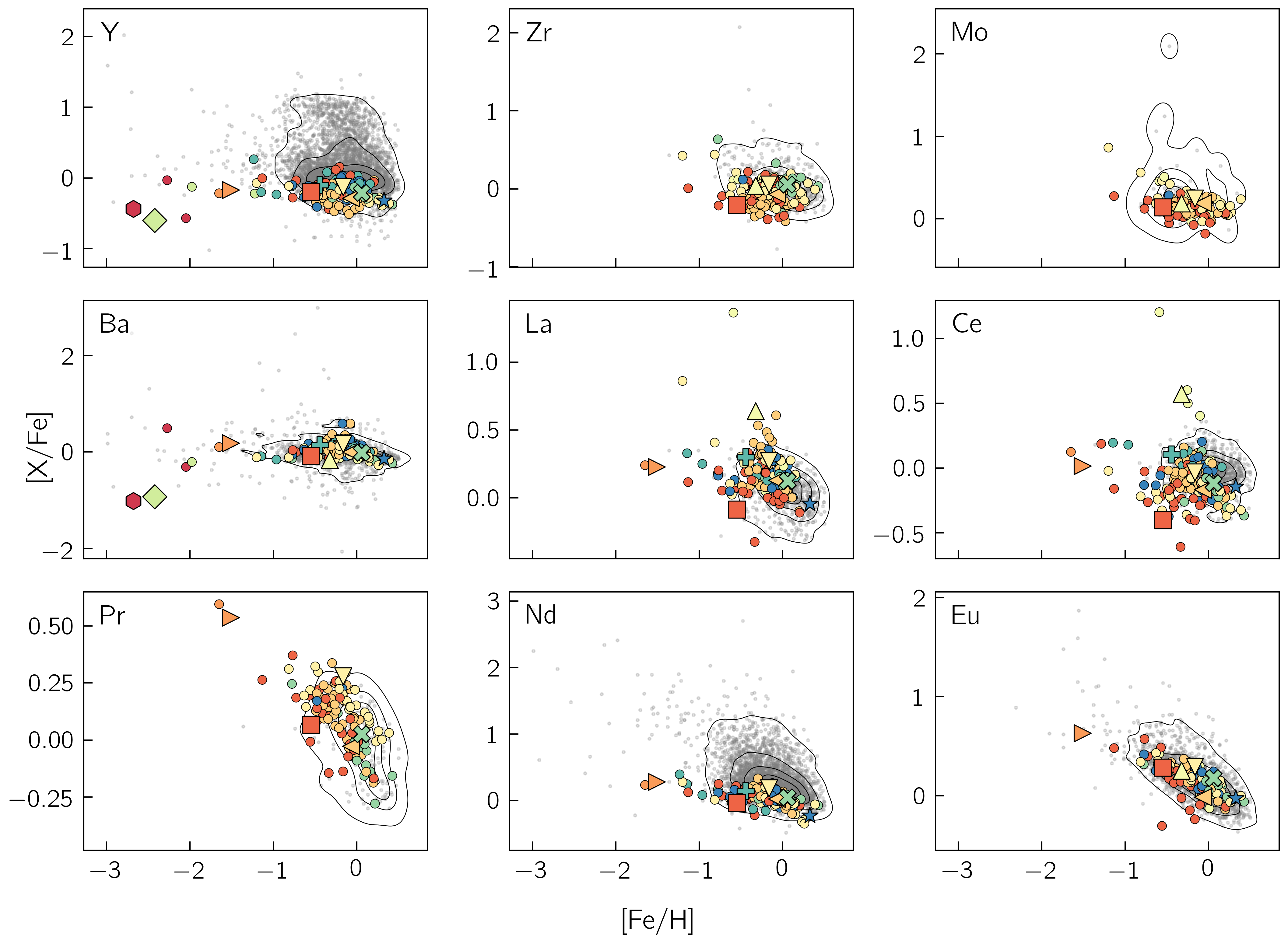}
\caption{Various [X/Fe]–[Fe/H] trends for the n-capture elements measured in this work. The GBS stars are color coded according to their group membership as assigned by the clustering analysis. The RepGBSs have the same symbols as adopted in Fig.~\ref{fig:hr_kmeans}. The gray points and contour plot represent abundances from the Gaia-ESO Survey \citep{2023A&A...676A.129H}.}
   \label{fig:ges}
\end{figure*}

The Sun was assigned to Group~III (\classrepIII) in our clustering procedure. Accordingly, the lines used to derive the final abundances follow the selection of \hiprepIII. The comparison of our results with the works mentioned above is shown in Fig.~\ref{fig:solar_delta}, while all the values can be seen in Table~\ref{Tab:solar}.

\ion{Y}{II}, as in the literature comparison outlined in Sect.~\ref{sect:literature_comp}, shows a notable difference in the solar case of 0.23 dex. To better understand this discrepancy, we refer to \cite{2015AGrevesse}, who present a detailed 3D-non-LTE analysis of n-capture elements building on \cite{2010Grevesse}, adopted by AS09. The difference may be attributed to the adopted spectral lines and oscillator strengths ($\log gf$). In particular, AS09 primarily uses $\log gf$ values from \cite{1982Hannaford}, with the exception of three Y lines not considered here, for which \cite{2011Biemont} is adopted. In our analysis, we instead rely entirely on the $\log gf$ values from \cite{2011Biemont}, as compiled by \citet{Heiter21}. However, it remains difficult to disentangle the effects of different $\log gf$ sources from possible 3D-non-LTE effects. Discrepancies of similar order of magnitude ( $\sim 0.2$ dex) also affect \ion{Ce}{II} and \ion{Pr}{II}. However, differences are observed not only between our results and the literature, but also between the two literature values themselves, which can again be related to the different transitions and $\log gf$ values adopted \citep[\ion{Pr}{II} values are reported in][]{2015AGrevesse}.

Although literature studies typically rely on \ion{Zr}{II} lines to infer the final abundance of this element, the detailed analysis of \cite{2015AGrevesse} is based on a significantly larger number of lines (ten in total), compared to the single \ion{Zr}{II} line at 511.227~nm used in our work. For this reason, we report the Zr abundance differences by comparing the literature values with the average of our results obtained from both ionization stages, while acknowledging that \ion{Zr}{I} is not the recommended ionization stage in the other two literature studies. This different approach may contribute to the $0.09$ dex underestimation of the Zr abundance observed in our results. Given these limitations, our comparison should be interpreted with caution. Nevertheless, it may suggest that Zr abundances are more reliably constrained using ionized lines in the bluer spectral region ($\lesssim 480$ nm). 

Abundances for \ion{Ba}{II} and \ion{Nd}{II} are consistent with the literature within uncertainties, whereas for \ion{La}{II} our results are higher by $\sim 0.1$ dex. Looking more closely at the literature, the \ion{La}{II} atomic data adopted by GR07 and AS09 are taken from \citep[][see their Table 4]{2001Lawler}, where 14 lines are listed, the majority of them lying in a bluer wavelength region than ours, which can partially explain the origin of this discrepancy. Nevertheless, our solar La abundance is in good agreement with the earlier determination of $\mathrm{A}_{\odot}(\ion{La}{II}) = 1.22$ reported by \citet{1998Grevesse}. Finally, for Eu, since only one line is used in our analysis, the offsets of 0.05 dex may be attributed to this limitation.

\subsection{Implications for Galactic studies}\label{sect:final_abund}

Our abundances for all GBSv3 are shown in Fig.~\ref{fig:ges}, together with the same literature sample for the solar neighborhood used in PVIII. The comparison stars are taken from the GES \citep{2023A&A...676A.129H}, selected as type “GE\_MW,” observed with UVES, and scaled to the solar abundances of \citet{grevesse07}, as adopted in this work. The literature values are shown in gray with contours, and our results are shown as circles with colors according to the clustering illustrated in Fig.~\ref{fig:hr_kmeans}. 

The [X/Fe] versus [Fe/H] trends for the GBS sample generally follow those observed in literature \citep[see also][]{2018Prantzos,2021A&A...649A.126T} and show good agreement with the GES data. For some elements, such as Ce and Pr, our high resolution spectra and careful analysis allowed us to extend the measurements to regimes that are slightly more metal poor. Notably, elements classified as light \textit{s}-process species (Y and Zr), exhibit a relatively large scatter (average up to $\sim 0.2$–$0.3$ dex), although smaller than that observed in the GES sample ($\gtrsim 0.3$ dex) of \cite{2023A&A...676A.129H}. 

Moving to the heavy \textit{s}-process elements measured in our sample, which are predominantly produced via production channels operating in low- and intermediate-mass asymptotic giant branch (AGB) stars \citep{2016Karakas}, we observe an increasing dispersion with decreasing metallicity ([Fe/H] $\lesssim -1.0$). For Ba, there is a hint of a trend similar to \cite{2021A&A...649A.126T}, in which the abundance increases up to roughly [Fe/H] $\sim -0.5$ dex and then declines, likely due to delayed $s$-process enrichment \citep{2018Prantzos}. Ce and La show dispersions comparable to the comparison sample, but larger than that of Ba, making it less straightforward to delineate a clear trend. However, La shows a slight indication of bimodality. While part of this spread for these elements may arise from measurement uncertainties and the known increase in dispersion at lower metallicity, it may also reflect production through both the weak and main \textit{s}-process in AGB stars, combined with possible contributions from additional rapid (\textit{r})-process elements \citep{2004Travaglio,2018Spite,2020Prantzos}. This combination would contribute to the overall increased dispersion. Finally, both La and Ce include the same star belonging to group IV (\classrepIV, shown in yellow), which is highly enhanced in both elements ([X/Fe] $\gtrsim 1.0$ dex).

Mo, which can be produced by both the $s$-process and the $r$-process \citep{2014ApJ...787...10B}, shows a distribution very similar to that found in the solar neighborhood. It exhibits a tight scatter over the explored metallicity range (comparable to that reported in works such as \citealt{Mishenina2019}) and a lack of more metal-poor targets in both samples, reflecting the difficulty of measuring this species in metal-poor stars. Pr and Nd, which also have a more mixed \textit{r}- and \textit{s}-process origin than Ba, La, and Ce \citep{2014ApJ...787...10B}, display, especially for the GBS stars, tighter trends with metallicity, showing a stronger decline indicative of a more uniform production channel likely dominated by the \textit{r}-process. [Pr/Fe] is enhanced at [Fe/H] $\lesssim -0.2$ dex, reflecting a stronger \textit{r}-process contribution compared to Ba, La, or Ce. The two stars in the Pr panel with [Pr/Fe] $> 0.5$ dex, belonging to group VIII (\classrepVIII), include the RepGBS \hiprepVIII, while the other is one of the more metal-poor stars for which we can measure the abundance of \ion{Pr}{II}. Furthermore, the trend [Pr/Fe] resembles that of [Eu/Fe], while that of [Nd/Fe] is flatter due to a larger s-process contribution, indicating that Pr has a relatively higher \textit{r-}process fraction. 

Eu, a nearly pure \textit{r}-process element \citep{1991Cowan,2018Cote}, shows a clear declining trend with metallicity, consistent with its origin in rapid and prompt events associated with short timescale enrichment, such as core-collapse supernovae or more exotic events \citep[neutron-star mergers, magneto-rotational supernovae, e.g.,][]{2014MNRAS.438.2177M,2015Cescutti,2020Kobayashi}.  
In summary, the last three panels of Fig.~\ref{fig:ges}, where the declines with metallicity are most evident, reflect Galactic enrichment from sources that enriched the Galaxy at early times and on short timescales, in contrast to lighter elements that experience delayed enrichment from the \textit{s}-process in low- and intermediate-mass AGB stars.

\section{Summary and conclusions}\label{sect:5}
Building on the analysis framework established in previous GBS studies, which already ensured homogeneity and accuracy in atmospheric parameters and in $\alpha$- and Fe-peak abundances, in this work we have presented a homogeneous determination of nine n-capture element abundances, namely Y, Zr, Mo, Ba, La, Ce, Pr, Nd, and Eu. Line-by-line abundances were derived via spectral synthesis, following the methodology of PVIII and using the TURBOSPECTRUM code. We started from a detailed line selection tailored to different stellar groups defined around the same ten RepGBSs adopted in PVIII and constructed through a clustering approach based on spectral features in Sect.~\ref{sect:2} and Sect.~\ref{sect:3}. In this way, we accounted for the diversity in stellar parameters and optimized the reliability of abundance determinations across the full sample. A thorough line assessment addressing the intrinsic challenges associated with n-capture elements (including the limited number of usable spectral lines, blending effects, and line weakness) was carried out for the RepGBS, for which a careful comparison with the literature was also addressed (Sect.~\ref{sect:literature_comp}). From this analysis, we have identified the difficulties in deriving heavy element abundances for metal-poor stars in the wavelength range covered by our spectra. These stars are also those most underrepresented in both our results and the literature (groups II, IV, and IX). 
These are likewise the stars that show the largest internal scatter and measurement uncertainties because of the few and weak lines available. Nevertheless, for the majority of elements and stars, good consistency among measurements from different spectrographs (shown in the plots of Sect.~\ref{sect:3}) and with literature values (Fig.~\ref{fig:litgbs}) demonstrate the robustness of our analysis. 

Surprisingly \ion{Y}{II}, despite having more lines available for its final determination, was one of the elements showing the largest differences with respect to the literature,  including in the comparison with the solar literature values. Nevertheless, the compilation of data from different sources, implying different adopted atmospheric parameters, methods, and line selections, complicates the identification of the underlying reasons for these discrepancies. This highlights the importance of having homogeneous reference catalogs, such as this one, that provide abundances for complex chemical elements, such as the n-capture species. It is important to have a solid understanding of these measurements because n-capture elements are key to the study of the stellar populations of our Galaxy \citep{2003Sneden} as well as to stellar evolution (e.g., for investigating binarity and more evolved phases, \citealt{2023Escorza}).

Finally, in Sect.~\ref{sect:4} we explored the robustness of our line selection and its extension to the entire GBSv3 sample. The number of measurements per element varies across the sample, with coverage ranging from 32\% for \ion{Zr}{II} up to 86\%, while most elements are measured in approximately 60–85\% of the total sample. The typical uncertainties range between $\sim 0.1$ and $0.3$ dex, depending on the element and type of stars. 

This section concludes with a comparison between our [X/Fe] trends and those of Milky Way stars selected from the Gaia-ESO Survey.
A good overlap between the distributions of the two samples is observed in Fig.~\ref{fig:ges}, with our measurements extending the metallicity range for some elements. The expected Galactic trends are visible along with the known increase in scatter toward lower metallicities, reflecting the challenges associated with the low-[Fe/H] regime but also the stochasticity of n-capture chemical enrichment in the early history of the Milky Way assembly. 

This study represents a step forward in extending the GBS series of works (particularly those of \cite{2015jofre} and \cite{2025Casamiquela}, which provided abundances of iron-peak and $\alpha$-capture elements for the GBS) to the release of reference abundances of new elements, namely the n-capture ones. These elements are key in Galactic archaeology studies but are challenging to measure and thus have lacked (until now) a homogeneous reference for a wide range of stars. Our measurements provide a valuable benchmark for validating and calibrating abundance pipelines as well as for understanding the differences between surveys and catalogs \citep{2023Heged,2025Buder}. In this context, they offer a well-characterized reference for elements that remain challenging to measure in large automated analyses due to line weakness, blending, or saturation effects \citep{2021Karinkuzhi,2023Kordopatis,2025Manea}. As such, this sample serves as a crucial testbed for improving the reliability and homogeneity of abundance determinations in current and future spectroscopic surveys such as GALAH, WEAVE, 4MOST, and the After Sloan 5.

\section*{Data availability}
The spectra and final abundance results can be found in our dedicated website of the Gaia benchmark stars \footnote{\url{https://www.blancocuaresma.com/s/benchmarkstars}}. We furthermore provide the final abundance of all elements studied here for all stars;  the line-by-line and instrument-by-instrument equivalent widths, abundances, and uncertainties for all elements and stars; and the atomic information of all the lines used here through the CDS. 

\begin{acknowledgements}
The preparation of this work has made extensive use of Topcat \citep{taylor2005topcat}, of the Simbad and VizieR databases at CDS, Strasbourg, France. Additionally, this research has made use of the Astrophysics Data System, funded by NASA under Cooperative Agreement 80NSSC21M0056.
We warmly thank Carme Jordi for her contribution to the observations of the GBS and Thomas Nordlander for the fruitful discussions. This work was financially supported by FONDECYT Regular grant Number 1231057. SV and AE acknowledges financial support from “La Caixa” Foundation (ID 100010434) with fellowship code LCF/BQ/PI23/11970031. 
SV also acknowledges financial support from the Spanish Ministry of Science, Innovation and Universities (MICIU) project PID2023-149982NB-I00. I.H.A., and P.J. acknowledge financial support from FONDECYT Regular 1231057. C.A.G. acknowledges financial support from FONDECYT Regular 1262342. U.H. acknowledges support from the Swedish National Space Agency (SNSA/Rymdstyrelsen).

\end{acknowledgements}

\bibliographystyle{aa}
\bibliography{references}
\appendix

\section{Line selection summary}
Table~\ref{tab:lines} lists the spectral lines adopted for the n-capture elements across the RepGBS as discussed in Sect.~\ref{sect:3}.
\begin{table*}[t!] 
\centering
\small
\caption{Line selection adopted for all n-capture elements in the ten RepGBS sample.}
\label{tab:lines}
\begin{tiny}
\begin{tabular}{lllllllllll}
\hline
\textbf{Line [nm]} &  HIP~67927 & HIP~32851 & HIP~79672 & HIP~76976 & HIP~104214 & HIP~95362 & HIP~37826& HIP~92167 & HIP~69673 & HIP~68594 \\
& $\eta$~Boo & HD~49933 & 18~Sco & HD~140283 & 61~Cyg~A & HD~182736 & 
bet~Gem & HD~175305 & Arcturus & HD~122563 \\
 & \classrepI &\classrepII & \classrepIII & \classrepIV & \classrepV & \classrepVI & \classrepVII & \classrepVIII & \classrepIX & \classrepX \\
\hline
Y1\_563.013  &  x & x & x & x & x & x & x & x & x & x \\
Y2\_488.368  &  \textcolor{green}{$\boldsymbol{\checkmark}$} & \textcolor{green}{$\boldsymbol{\checkmark}$} & \textcolor{green}{$\boldsymbol{\checkmark}$} & \textcolor{green}{$\boldsymbol{\checkmark}$} & x & \textcolor{green}{$\boldsymbol{\checkmark}$} & \textcolor{green}{$\boldsymbol{\checkmark}$}  & \textcolor{green}{$\boldsymbol{\checkmark}$} & x & \textcolor{green}{$\boldsymbol{\checkmark}$} \\
Y2\_490.012  &  x & x & \textcolor{green}{$\boldsymbol{\checkmark}$} & \textcolor{green}{$\boldsymbol{\checkmark}$} & x & x & \textcolor{green}{$\boldsymbol{\checkmark}$} & \textcolor{green}{$\boldsymbol{\checkmark}$} & x & \textcolor{green}{$\boldsymbol{\checkmark}$} \\
Y2\_508.742  &  \textcolor{green}{$\boldsymbol{\checkmark}$} & \textcolor{green}{$\boldsymbol{\checkmark}$} & \textcolor{green}{$\boldsymbol{\checkmark}$} & \textcolor{green}{$\boldsymbol{\checkmark}$} & x & \textcolor{green}{$\boldsymbol{\checkmark}$} & \textcolor{green}{$\boldsymbol{\checkmark}$} & \textcolor{green}{$\boldsymbol{\checkmark}$} & x & \textcolor{green}{$\boldsymbol{\checkmark}$} \\
Y2\_512.321*  &  x & \textcolor{green}{$\boldsymbol{\checkmark}$} & x & x & x & x & x & \textcolor{green}{$\boldsymbol{\checkmark}$} & x & x \\
Y2\_520.041  &  x & \textcolor{green}{$\boldsymbol{\checkmark}$} & \textcolor{green}{$\boldsymbol{\checkmark}$} & x & x & \textcolor{green}{$\boldsymbol{\checkmark}$} & \textcolor{green}{$\boldsymbol{\checkmark}$} & \textcolor{green}{$\boldsymbol{\checkmark}$} & \textcolor{green}{$\boldsymbol{\checkmark}$} & \textcolor{green}{$\boldsymbol{\checkmark}$} \\
Y2\_520.572  &  x & x & \textcolor{green}{$\boldsymbol{\checkmark}$} & x & x & x & \textcolor{green}{$\boldsymbol{\checkmark}$} & \textcolor{green}{$\boldsymbol{\checkmark}$} & x & \textcolor{green}{$\boldsymbol{\checkmark}$} \\
Y2\_554.601  & \textcolor{green}{$\boldsymbol{\checkmark}$} & x & x & x & x & \textcolor{green}{$\boldsymbol{\checkmark}$} & \textcolor{green}{$\boldsymbol{\checkmark}$} & x & \textcolor{green}{$\boldsymbol{\checkmark}$} & x \\
\hline
Zr1\_480.587  & x & x & x & x & x & x& \textcolor{green}{$\boldsymbol{\checkmark}$} & x & \textcolor{green}{$\boldsymbol{\checkmark}$} & x \\
Zr1\_480.947 & x & x & x & x & x & x & x & x & \textcolor{green}{$\boldsymbol{\checkmark}$} & x \\
Zr1\_481.504 & x & x & x & x & x & \textcolor{green}{$\boldsymbol{\checkmark}$} & \textcolor{green}{$\boldsymbol{\checkmark}$} & x & \textcolor{green}{$\boldsymbol{\checkmark}$} & x \\
Zr1\_481.563 & x & x & x & x & x & x & \textcolor{green}{$\boldsymbol{\checkmark}$} & x & \textcolor{green}{$\boldsymbol{\checkmark}$} & x \\
Zr1\_482.804 & x & x & x & x & x & x & \textcolor{green}{$\boldsymbol{\checkmark}$} & x & \textcolor{green}{$\boldsymbol{\checkmark}$} & x \\
Zr1\_507.825$\dagger$ & x & x & x & x & x & x & x & x & x & x \\
Zr1\_612.744 & x & x & \textcolor{green}{$\boldsymbol{\checkmark}$} & x & \textcolor{green}{$\boldsymbol{\checkmark}$} & \textcolor{green}{$\boldsymbol{\checkmark}$} & \textcolor{green}{$\boldsymbol{\checkmark}$} & x & \textcolor{green}{$\boldsymbol{\checkmark}$} & x \\
Zr1\_613.455 & x & x & \textcolor{green}{$\boldsymbol{\checkmark}$} & x & \textcolor{green}{$\boldsymbol{\checkmark}$} & \textcolor{green}{$\boldsymbol{\checkmark}$} & \textcolor{green}{$\boldsymbol{\checkmark}$} & x & \textcolor{green}{$\boldsymbol{\checkmark}$} & x \\
Zr2\_511.227 & \textcolor{green}{$\boldsymbol{\checkmark}$} & \textcolor{green}{$\boldsymbol{\checkmark}$} & \textcolor{green}{$\boldsymbol{\checkmark}$} & x & x & \textcolor{green}{$\boldsymbol{\checkmark}$} & x & \textcolor{green}{$\boldsymbol{\checkmark}$} & x & x \\
\hline
Mo1\_553.303* & x & x & x & x & \textcolor{green}{$\boldsymbol{\checkmark}$} & x & x& x & x & x \\
Mo1\_557.044 & x & x & x & x & x & x & \textcolor{green}{$\boldsymbol{\checkmark}$} & x & \textcolor{green}{$\boldsymbol{\checkmark}$} & x \\
Mo1\_603.064 & x & x & x & x & \textcolor{green}{$\boldsymbol{\checkmark}$} & \textcolor{green}{$\boldsymbol{\checkmark}$} & \textcolor{green}{$\boldsymbol{\checkmark}$} & x & \textcolor{green}{$\boldsymbol{\checkmark}$} & x \\
\hline
Ba2\_585.367 & \textcolor{green}{$\boldsymbol{\checkmark}$} & \textcolor{green}{$\boldsymbol{\checkmark}$} & \textcolor{green}{$\boldsymbol{\checkmark}$} & x & \textcolor{green}{$\boldsymbol{\checkmark}$} & \textcolor{green}{$\boldsymbol{\checkmark}$} & \textcolor{green}{$\boldsymbol{\checkmark}$} & \textcolor{green}{$\boldsymbol{\checkmark}$} & \textcolor{green}{$\boldsymbol{\checkmark}$} &  \textcolor{green}{$\boldsymbol{\checkmark}$}\\
Ba2\_614.171 & x & x & x & \textcolor{green}{$\boldsymbol{\checkmark}$} & x &  x & x & \textcolor{green}{$\boldsymbol{\checkmark}$} &x  & \textcolor{green}{$\boldsymbol{\checkmark}$} \\
Ba2\_649.690 & x & x & x & x & \textcolor{green}{$\boldsymbol{\checkmark}$} & x & x & x & x &\textcolor{green}{$\boldsymbol{\checkmark}$} \\
\hline
La2\_480.403 & x & x& x & x & x & x& x & x&  \textcolor{green}{$\boldsymbol{\checkmark}$} & x \\ 
La2\_511.456$\dagger$ & \textcolor{green}{$\boldsymbol{\checkmark}$} & \textcolor{green}{$\boldsymbol{\checkmark}$} & \textcolor{green}{$\boldsymbol{\checkmark}$} & x & x & \textcolor{green}{$\boldsymbol{\checkmark}$} & \textcolor{green}{$\boldsymbol{\checkmark}$} & \textcolor{green}{$\boldsymbol{\checkmark}$} &  x& x \\ 
La2\_529.082$\dagger$ & \textcolor{green}{$\boldsymbol{\checkmark}$} & x& x & x & x & x& x &\textcolor{green}{$\boldsymbol{\checkmark}$}&  \textcolor{green}{$\boldsymbol{\checkmark}$} & x \\ 
La2\_632.038$\dagger$ & x & x& x & x & \textcolor{green}{$\boldsymbol{\checkmark}$}& x& x & x&  \textcolor{green}{$\boldsymbol{\checkmark}$} & x \\ 
La2\_639.046 & x & x& x & x & \textcolor{green}{$\boldsymbol{\checkmark}$} & x& x & \textcolor{green}{$\boldsymbol{\checkmark}$}&  \textcolor{green}{$\boldsymbol{\checkmark}$} & x \\ 
\hline
Ce2\_527.423 & \textcolor{green}{$\boldsymbol{\checkmark}$} & \textcolor{green}{$\boldsymbol{\checkmark}$} & \textcolor{green}{$\boldsymbol{\checkmark}$} & x & x & \textcolor{green}{$\boldsymbol{\checkmark}$} & \textcolor{green}{$\boldsymbol{\checkmark}$} & \textcolor{green}{$\boldsymbol{\checkmark}$} &  \textcolor{green}{$\boldsymbol{\checkmark}$}& x \\
Ce2\_533.056 & \textcolor{green}{$\boldsymbol{\checkmark}$} & \textcolor{green}{$\boldsymbol{\checkmark}$} & \textcolor{green}{$\boldsymbol{\checkmark}$} & x & x & x & \textcolor{green}{$\boldsymbol{\checkmark}$} & \textcolor{green}{$\boldsymbol{\checkmark}$} & x&x \\
Ce2\_561.025$\dagger$ &x & x& x & x & \textcolor{green}{$\boldsymbol{\checkmark}$} & x & \textcolor{green}{$\boldsymbol{\checkmark}$}  & x & x&x \\
Ce2\_604.337& x & x & \textcolor{green}{$\boldsymbol{\checkmark}$} & x & x & \textcolor{green}{$\boldsymbol{\checkmark}$} & x & x & \textcolor{green}{$\boldsymbol{\checkmark}$}&x \\	
\hline
Pr2\_513.514$\dagger$& x & x & x & x & x & x & x & \textcolor{green}{$\boldsymbol{\checkmark}$} & x & x \\
Pr2\_525.974& x & x & \textcolor{green}{$\boldsymbol{\checkmark}$} & x & x & \textcolor{green}{$\boldsymbol{\checkmark}$} & \textcolor{green}{$\boldsymbol{\checkmark}$} & \textcolor{green}{$\boldsymbol{\checkmark}$} &  \textcolor{green}{$\boldsymbol{\checkmark}$}& x \\
Pr2\_532.277& x & x & x & x & x & \textcolor{green}{$\boldsymbol{\checkmark}$} & \textcolor{green}{$\boldsymbol{\checkmark}$} & \textcolor{green}{$\boldsymbol{\checkmark}$} &  \textcolor{green}{$\boldsymbol{\checkmark}$}& x \\
\hline
Nd2\_513.059*& \textcolor{green}{$\boldsymbol{\checkmark}$} & \textcolor{green}{$\boldsymbol{\checkmark}$} & \textcolor{green}{$\boldsymbol{\checkmark}$} & x & x & \textcolor{green}{$\boldsymbol{\checkmark}$} & \textcolor{green}{$\boldsymbol{\checkmark}$} & \textcolor{green}{$\boldsymbol{\checkmark}$}& x & x \\
Nd2\_518.117$\dagger$ & x & x & x & x & x & x & x & x &  x & x \\
Nd2\_521.565$\dagger$& x & x & x & x & x & x & x & x &  x & x \\
Nd2\_529.316* & \textcolor{green}{$\boldsymbol{\checkmark}$} & \textcolor{green}{$\boldsymbol{\checkmark}$} & \textcolor{green}{$\boldsymbol{\checkmark}$} & x & x & x & x & \textcolor{green}{$\boldsymbol{\checkmark}$} & \textcolor{green}{$\boldsymbol{\checkmark}$}& x \\
Nd2\_531.981 & \textcolor{green}{$\boldsymbol{\checkmark}$} & x & \textcolor{green}{$\boldsymbol{\checkmark}$} & x & x & \textcolor{green}{$\boldsymbol{\checkmark}$} & \textcolor{green}{$\boldsymbol{\checkmark}$} & \textcolor{green}{$\boldsymbol{\checkmark}$} &  \textcolor{green}{$\boldsymbol{\checkmark}$}& x \\
Nd2\_568.852$\dagger$ & x & x & x & x & x & x & \textcolor{green}{$\boldsymbol{\checkmark}$} & \textcolor{green}{$\boldsymbol{\checkmark}$} & \textcolor{green}{$\boldsymbol{\checkmark}$} & x \\
\hline
Eu2\_664.510 & \textcolor{green}{$\boldsymbol{\checkmark}$} & x & \textcolor{green}{$\boldsymbol{\checkmark}$} & x & \textcolor{green}{$\boldsymbol{\checkmark}$} & \textcolor{green}{$\boldsymbol{\checkmark}$} & \textcolor{green}{$\boldsymbol{\checkmark}$} & \textcolor{green}{$\boldsymbol{\checkmark}$} & \textcolor{green}{$\boldsymbol{\checkmark}$} & x \\
\hline
\end{tabular}%
\end{tiny}
\tablefoot{ Symbols indicate: (*) \texttt{synflag = N}, and ($\dagger$) flags not available. Otherwise, the flags are either \texttt{Y} or \texttt{U}. The HIP identifiers, together with the common (catalog) names of the stars, are reported. The labels for the stellar groups follow the nomenclature defined in Table\ref{tab:repgb_nomenclature}. }
\tablebib{For oscillator strengths: \citet{1992AA...255..457D, 2007PhyS...76..577L, BBEHL, BGHL, CB, HLSC, ILW, K06, LBS, LNAJ, LSCI, LWHS, MC, WBb}; for data for line broadening by neutral hydrogen collisions: \citet{K11} for Y, \citet{BPM} for Zr, Mo, Ba.}
\end{table*}

\section{RepGBS nomenclature}\label{appendix:other}
A new nomenclature for the ten representative stars (RepGBS), adopted from PVIII (see their Table 1), is introduced in Table~\ref{tab:repgb_nomenclature}. It is designed to reflect both the spectral type and the metallicity of each star (as derived in PVIII and reported in the last column of the table) and is used consistently throughout this work.

\begin{table}[H]
\centering
\caption{Nomenclature of the ten RepGBSs used in this work.}
\label{tab:repgb_nomenclature}

\begin{tabular}{|l|rrrr|}

\hline
\textbf{RepGBS} & \textbf{Spectral type}  & \textbf{Group} & \textbf{\#} & \textbf{[Fe/H]} \\
\hline
\hiprepI        & \typerepI         & \classrepI & I & 0.32 \\
\hiprepII           & \typerepII        & \classrepII & II & -0.45\\
\hiprepIII          & \typerepIII   & \classrepIII & III & 0.06 \\
\hiprepIV         & \typerepIV  & \classrepIV & IV & -2.42\\
\hiprepV  & \typerepV & \classrepV & V& -0.33\\
\hiprepVI & \typerepVI & \classrepVI &VI& -0.17\\
\hiprepVII & \typerepVII & \classrepVII & VII& -0.07\\
\hiprepVIII & \typerepVIII  & \classrepVIII & VIII & -1.51\\
\hiprepIX         & \typerepIX  & \classrepIX & IX & -0.55 \\
\hiprepX    & \typerepX & \classrepX & X & -2.68\\
\hline

\end{tabular}
\tablefoot{The nomenclature encodes both the spectral type and the metallicity values. The group to which each RepGBS is assigned is shown in the fourth column, while the last column lists the metallicity values inferred in PVIII.}
\end{table}

\section{Elements analyzed but excluded from final abundances}\label{appendix:other}
\subsection{Strontium}

Strontium (Sr) abundances were initially measured using the \ion{Sr}{I} line at 496.226~nm. However, this line is too weak in \classrepI\ and \classrepII\ (\hiprepI, \hiprepII) and in metal-poor stars (\hiprepIV, \hiprepX), resulting in unreliable measurements. For cooler, more metal-rich stars, the line could be measured, but instrument-to-instrument variations and continuum placement still affected the precision \citep[e.g.,][]{Heiter21,2014Barbuy,2018MKarinkuzhi,2025Casamiquela}. The Sr1\_496.226~nm line is flagged as {\tt Y/N} in the GES linelist, indicating reliable atomic data but less accurate synthesis in the Sun and Arcturus. Attempts to use additional Sr lines from the literature \citep[481.18, 640.84, 650.39, and 679.10~nm;][]{2014Barbuy,2018MKarinkuzhi,2021Karinkuzhi} did not yield a better result. Thus, we do not provide Sr abundances, noting that better and more detailed Sr lines lie at wavelengths outside our observed range  \citep{2021Karinkuzhi}.

\subsection{Samarium}

As noted in previous studies, only a few Sm lines in the optical wavelength range can in principle be used reliably, due to the strong impact of hyperfine structure (HFS) and isotope ratios \citep{2007Lundqvist,Heiter21}. For this reason, chemical abundance studies of Sm have traditionally focused on bluer spectral regions \citep{2008Roederer,2025Lombardo}.

We attempted to measure \ion{Sm}{II} abundances from the available transitions, but issues such as line weakness, blending, and continuum placement uncertainties prevented reliable determinations. Only a few lines could be considered in individual stars. For example, in \hiprepI\ two of four transitions were tentatively inspected, in \hiprepVII\ only the Sm2\_497.217~nm line could be considered, and in \hiprepIII\ (\classrepIII) none of the lines could be reliably used. Even lines that could be synthesized accounting for HFS, such as the 483.462~nm transition, were insufficient to provide robust results. Indeed, our comparison with the literature showed very large discrepancies confirming that our measurements were too uncertain. 

\section{Examples of challenging spectral lines}\label{appendix:lines}
We present some examples of line profiles that despite being flagged as suitable according to the REW criterion and line-by-line abundance dispersions were either discarded or treated with caution when deriving the final abundances.
\begin{figure}[t]
    \centering    \includegraphics[width=0.99\linewidth]{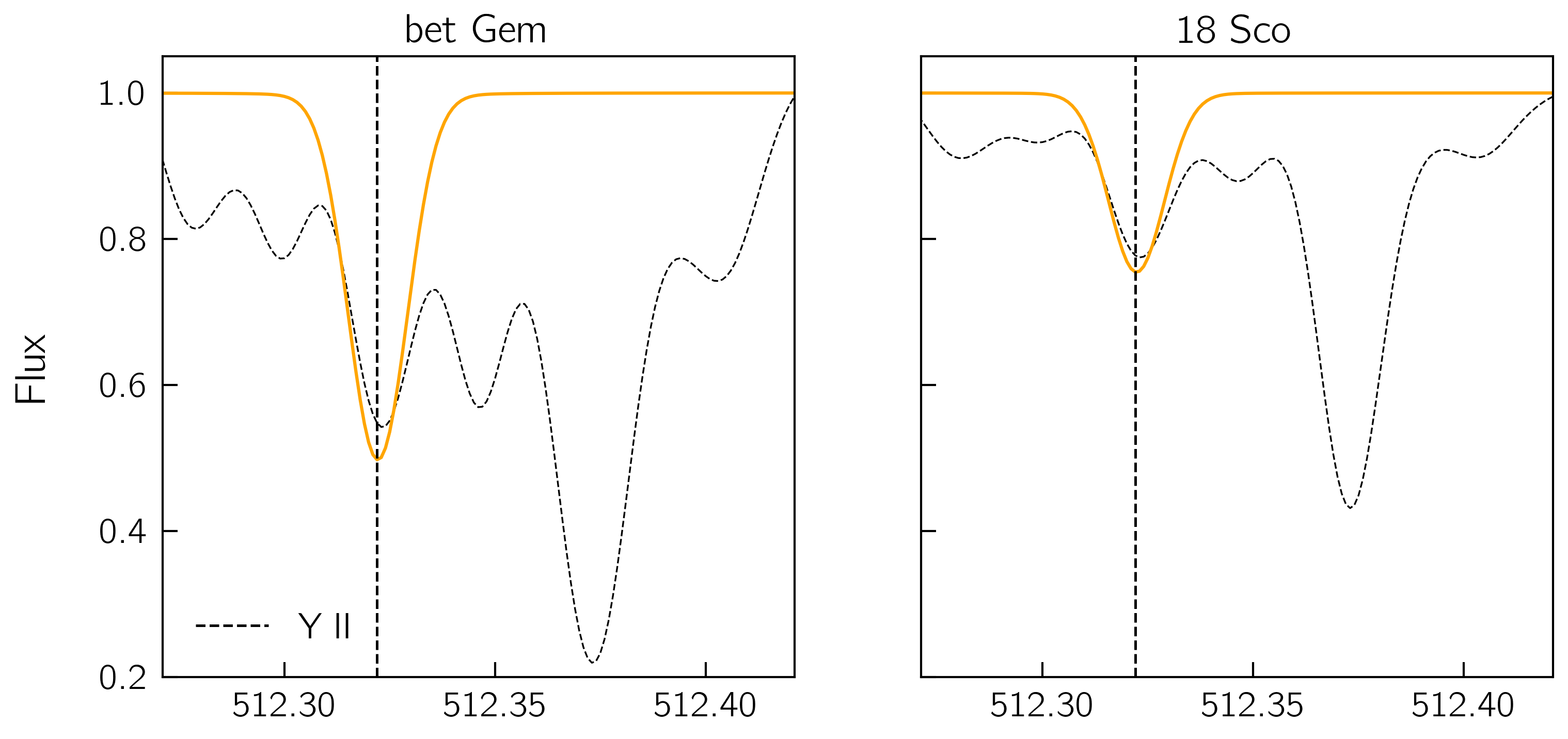}
\includegraphics[width=0.99\linewidth]{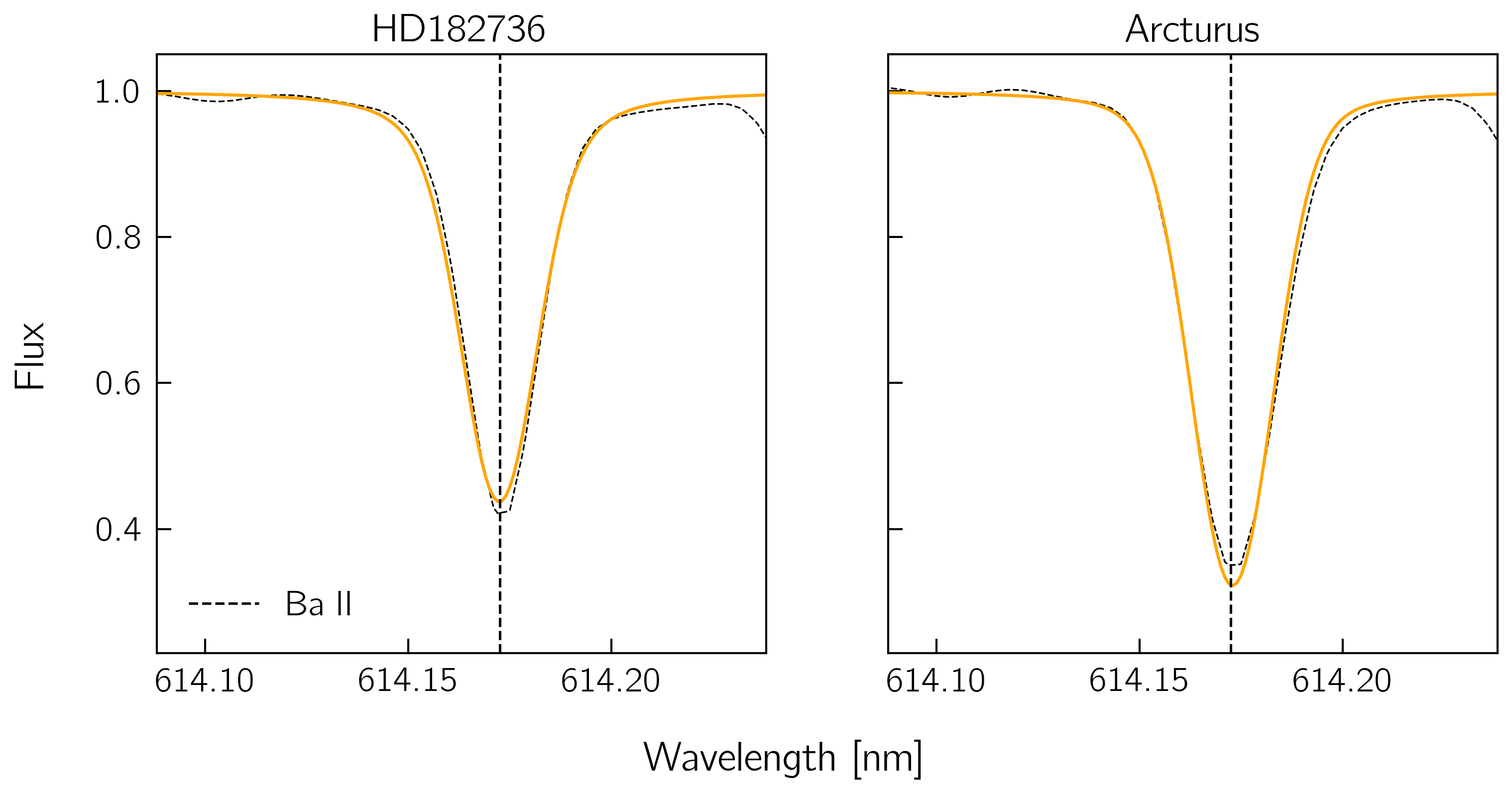}
\includegraphics[width=0.99\linewidth]{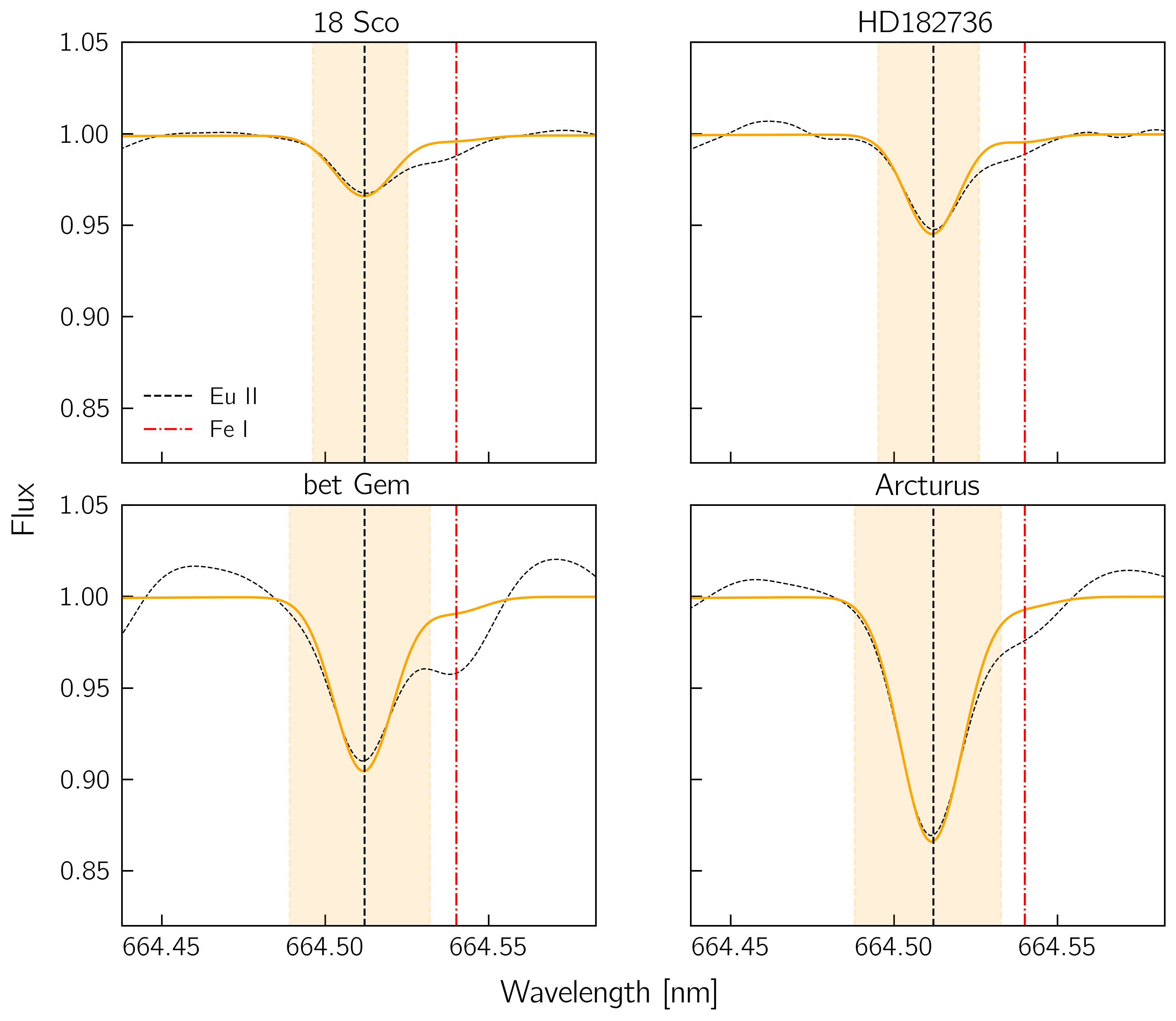}
    \caption{The four top panels show spectra of four RepGBS observed with HARPS and NARVAL for the \ion{Y}{II} line at 512.321~nm and the \ion{Ba}{II} line at 614.171~nm, respectively. The observed spectrum is shown as a black dashed line, while the single-line best-fitting synthetic spectrum is shown as a continuous orange line. The centroids of the four transitions are marked with dashed black lines. The bottom four panels illustrate example spectra of four RepGBS observed with ESPaDOnS. The observed spectrum is shown as a black dashed line, while the best-fitting synthetic spectrum of the single \ion{Eu}{II} line used in this work is plotted as a continuous orange line. The orange shaded region indicates the wavelength interval used to derive the abundance from the \ion{Eu}{II} line, whose central wavelength is marked by the black dashed vertical line at 664.51~nm. The nearby \ion{Fe}{I} feature, with central wavelength shown as a red dash-dotted line, lies outside this fitting window and is therefore not included in the spectral synthesis used to derive the Eu abundance.}  \label{fig:yba_panels}
\end{figure}

Y and Ba, shown in Fig.~\ref{fig:yba_panels}, provide examples of two lines marked with black circles in Fig.~\ref{fig:lines_new} (i.e., good measurements) for some RepGBS that were nevertheless rejected. The Y2\_512.321 line, flagged as \texttt{N} in the GES line list, was excluded for most RepGBS because of continuum-placement difficulties caused by nearby molecular features (TiO bands). This is illustrated for two examples, namely \hiprepIII\ and \hiprepVII. The Ba2\_614.171 line, reported for the giant stars \hiprepVI\ and \hiprepIX, also shows measurements comparable to the other transitions considered in this work, but in this case it is affected by saturation effects, which are evident in the line cores, where the LTE synthetic fit is not able to reproduce the observed profile.

A different situation is represented by the \ion{Eu}{II} line, which is affected by blending with nearby \ion{Fe}{I} features. In these cases, the blend can be reliably accounted for through careful spectral synthesis. To minimize its impact on the abundance determination, we adopt a restricted spectral window centered on the Eu feature during the fitting procedure, as illustrated by the ESPaDOnS spectra of the metal-rich dwarf \hiprepIII\ and the three giant stars shown at the bottom of Fig.~\ref{fig:yba_panels}.

\section{Literature compilation for RepGBS}\label{app:lit}
We summarize below the reference samples adopted from the literature to build the comparison described in Sect.~\ref{sect:literature_comp}. 
Table~\ref{tab:lit} includes the literature compilation for the abundances analyzed in this work used to compare our results in Fig.~\ref{fig:litgbs}.  In the following we summarize the key characteristics of the literature results we use in our work. 

\begin{table*}[t!]
\caption{Literature compilation of abundances for RepGBS.}
\centering
\begin{tabular}{|l|rrrrrrrrrrrr|}
\hline
  \multicolumn{1}{|c|}{star} &
  \multicolumn{1}{c}{Sr} &
  \multicolumn{1}{c}{Y} &
  \multicolumn{1}{c}{Zr1} &
 \multicolumn{1}{c}{Zr2} &
  \multicolumn{1}{c}{Mo} &
  \multicolumn{1}{c}{Ba} &
  \multicolumn{1}{c}{La} &
  \multicolumn{1}{c}{Ce} &
  \multicolumn{1}{c}{Pr} &
  \multicolumn{1}{c}{Nd} &
  \multicolumn{1}{c}{Sm} &
  \multicolumn{1}{c|}{Eu}\\ 
\hline 
  \hiprepI & 0.02 & -0.21 & & -0.26 &  & -0.24 & -0.03 & -0.20 & -0.21 & -0.28 & 0.01 & -0.07 
  \\ HIP 67927  & H14 & H23  & & H23  & &  H23 & H23 &  H23& L17 & H23 & L17 & H23\\ \hline
  \hiprepII & -0.08 & -0.04 & & 0.09 & & 0.35 & 0.16 & 0.23 &  & 0.27 & 0.44 & 0.11 \\ 
  HIP 32851 & R09 & R09 & & R09 &  & R09 & R09 & R09 &  & R09 & R09 & R09\\ \hline 
  \hiprepIII  & 0.04 & 0.03 & 0.05 & -0.07 &  0.02 & 0.06 & 0.06 & 0.08 & 0.06 & 0.10 & -0.06 & 0.13 \\ HIP 79672 & M14 & H23 & M14 & H23 & M14 & M14 & M14 & M14 & M14 & M14 & M14 & M14\\ \hline
 \hiprepIV & -0.18  & -0.40 & & -0.07 & -0.04 & -0.81 & $<-0.36$ & 0.18  &  & -0.45 &  & -0.28 \\ 
 HIP 76976    & S15 & S15 & & S15 & P20 & S15 & S15 & S15 &  & H23 & & S15 \\ \hline 
 \hiprepV & 0.66 &  &  -0.53 & &  & -0.08 & 0.79 & 1.52 & 1.5 & 1.84 & 2.33 & 0.39 \\ 
 HIP 104214 & L17 &  & L17  & &  & L17 &  L17& L17 & L17 & L17 &L17& L17 \\ \hline
  \hiprepVI & 0.08 & -0.05 & -0.02 & 0.04  & & 0.05 & 0.32 & -0.02 & 0.31 & 0.38 & 0.07 & 0.61\\
  HIP 95362& L17 & M13 & L17 &M13&  & M13 & L17 & M13 &  L17 & L17 & M13& L17 \\ \hline
  \hiprepVII & -0.02 & 0.10 &  0.06 &0.21  & 0.26 & 0.06 & 0.14 & 0.28 & 0.12 & 0.07 & -0.67 & 0.05 \\ 
  HIP 37826 & L05 & S25 & S25 & S25 & F22 & S25 &  S25 & S25 & S25 & S25 &L05 & S25 \\ \hline
  \hiprepVIII & -0.14 & -0.34 & & 0.2  & 0.11 & 0.1 & 0.08 & 0.07 &  & 0.29 & 0.16 & 0.4\\ 
  HIP 92167 & R14 & R14 & & R14 & R14  & R14 & R14 & R14 &  & R14 & R14& R14\\ \hline
  \hiprepIX & -0.05 & -0.11 & -0.12 & -0.03 & 0.17 & 0.02 & -0.09 & -0.23 & -0.09 & -0.05 & 0.12 & 0.25\\ 
  HIP 69673 & K18 & K18 & K18 & K18 & B25 & B25 & K18 & K18 & K18 & K18 & K18 & K18\\ \hline
  \hiprepX & -0.19 & -0.62 & & -0.2 &   & -1.17 & -1.03 & -0.93 &  & -1.02 & -0.53 & -0.84 
  \\ HIP 68594 &  R14& R14 & & R14 &  & R14 & R14 & R14 &  & R14 & R14& R14\\
\hline\end{tabular}
\tablefoot{Abundances in [X/Fe] using the Fe abundance adopted by each work.   H14: \cite{2014AJ....148...54H}; H23: \cite{2023A&A...676A.129H};  R09: \cite{2009A&A...506..203R}; M14: \cite{Melendez14}; S15: \cite{2015A&A...584A..86S}; P20: \cite{2020A&A...638A..64P}; M13: \cite{2013A&A...552A.128M}; L17: \cite{2017AJ....153...21L}; S25: \cite{2025A&A...701A.153S}; F22: \cite{2022A&A...666A.125F}; R14: \cite{2014AJ....147..136R}; K18: \cite{2018A&A...618A..32K}; B25: \cite{2025PASA...42...51B}; L05: \cite{2005AJ....129.1063L}}\label{tab:lit}
\end{table*}

\subsection{B25 - \cite{2025PASA...42...51B}}
This is the latest data release of GALAH. The pipeline uses SME at massive scales, uses MARCS atmospheric models and applies non-LTE corrections. The spectra is of lower resolution and wavelength coverage than the other references but has been designed to include spectral windows which include lines of several n-capture elements. We consider these results for \hiprepIX\ for Mo and Ba abundances only, but many more are available.

\subsection{F22 - \cite{2022A&A...666A.125F}}
This work reports Mo, Ce and Eu abundances of bulge and disk stars, which include \hiprepVII. They use UVES spectra for the bulge stars, and FIES, ESPaDOnS and NARVAL data for the disk stars. The analysis was performed using SME \cite{2017Piskunov}, the GES linelist, and MARCS atmosphere models. The stellar parameters are calibrated with the GBS (PI and PIII). To determine Mo, they use the 603 nm line.

\subsection{H14 - \cite{2014AJ....148...54H}}
This is the Hypatia catalog, which collects abundances of stars in the solar neighborhood from various published papers. Hypatia has an up-to-date website\footnote{\url{https://www.hypatiacatalog.com/hypatia}} with an interface to select elements, stars or samples. Crucially, the abundances can be queried choosing a specific solar abundance scale, which in our case we chose \cite{grevesse07} because this is the one used in our {\tt iSpec} analysis. 

When a given star has abundances reported in more than one source, the abundance provided in the catalog can be either the median or the mean of all reported values, and the error corresponds to the standard deviation of these measurements. While these values, which have no restriction of model atmosphere or line list account for the true spread in stellar abundances, it makes it hard for us to perform a direct comparison and understand the differences. Thus, we adopted the values from Hypatia only for Y and Ba of \hiprepV. 

\subsection{H23 - \cite{2023A&A...676A.129H}}
This is the final compilation of the GES Survey \citep[][GES]{2022A&A...666A.120G}. Similarly to the results coming from the Hypatia catalog, the GES combines results from a variety of methods and instrument setups. In this case, however, a unique set of model atmospheres, solar abundance scale, line list was used, which is the same as the one adopted in our work. Moreover, all methods used the same set of stellar parameters for the stars. The uncertainties of the GES therefore reflect the methodology spread, such as the choice of lines, continuum normalization, and radiative transfer code.  

Since the GES used the GBS as standards for the stellar parameter scale, some of the GBSv1 stars contain measurements of several n-capture elements such as \hiprepI. We comment that the wavelength range and the resolution of our spectra has been designed to match the GES UVES 580 setup.

\subsection{K18: \cite{2018A&A...618A..32K}}
This paper investigates the role of n-capture elements in binary stars and used \hiprepIX\ as one of the reference stars in their analysis, providing detailed abundance determination in their Table 2. The analysis for this star is performed using BACCHUS, which is based on 1D-LTE Turbospectrum synthesis, MARCS model atmospheres, HERMES spectra and \cite{2009Asplund} solar scale. Their line list is in Table B1.  

\subsection{L05/L17 - \cite{2005AJ....129.1063L, 2017AJ....153...21L}}
These works present a chemical analysis of about 1000 nearby stars, including \hiprepV\ and \hiprepVI.  
They use spectra from the McDonald Observatory which cover a wavelength range similar to ours. The linelist used in these works adopt oscillator strengths that are astrophysically calibrated with the solar abundances. The scale used comes from \cite{2015A&A...573A..26S} and \cite{2015AGrevesse}. For our comparison in Sect.~\ref{sect:literature_comp} we list in Tab.~\ref{tab:lit} values from this work as fillers, but we note that additional measurements of \ion{Y}{I}, \ion{Zr}{II}, \ion{Ba}{II}, \ion{La}{II}, \ion{Nd}{II}, \ion{Eu}{II} for \hiprepI, \hiprepIII\, and \hiprepIV\  are also reported in these papers.

\subsection{M14 - \cite{Melendez14}}

These are values measured differentially with respect to the Sun using UVES data. They measure the sun abundances too. Solar values are those of \cite{2009Asplund}. They used EW measured by ARES \citep{2007A&A...469..783S} and MOOG to derive abundances. The paper provides ultra high precision abundances for all the elements investigated in this work, including detailed information at a line-by-line basis. The parameters they use for \hiprepIII\ are $5823 \pm 6$K for $\mathrm{T}_{\mathrm{eff}}$,  $4.45 \pm 0.02$ for $\log g$  and 0.045 for [Fe/H]. They provided Mo abundances using the very blue lines at 315.8 and 319.3 nm, and Zr abundances using the \ion{Zr}{II} at 405, 420.8 and 444.2 nm lines. All of these lines lie outside the wavelength range of our library.

\subsection{M13 - \cite{2013A&A...552A.128M}}
This paper reports n-capture abundances of more than 250 stars in the Galactic disk, including \hiprepVI. Their results come from the analysis of ELODIE spectra, the temperature was derived using Balmer lines and the surface gravity considering the ionization balance. Abundances were derived considering Kurucz models and WIDTH9. In particular,  Eu abundances were derived using STARSP \citep{1996ASPC..108..198T}. All lines and atomic information used in this work are tabulated in their Table 2.  They use their own solar abundance scale, obtained after analyzing a spectrum from the Moon. Ba abundances have non-LTE corrections using MULTI. 

\subsection{P20 - \cite{2020A&A...638A..64P}}
This paper focuses on the determination of heavy elements of five metal-poor dwarfs, including \hiprepIV. To do so, they analyzed HST spectra, that is, UV lines. This is how they were able to obtain [Mo/Fe] abundances in these metal-poor stars. Notably, they considered six Mo lines in the UV. The abundance was estimated using LTE analysis with SYNTHE and Turbospectrum.    

\subsubsection{R14 - \cite{2014AJ....147..136R}}
This paper reports on the abundances of more than 300 metal poor stars, including \hiprepVIII\ and \hiprepX. The spectra were taken with MIKE, and the abundances are derived using EW and the ionization/excitation balance approach, with the help of isochrones and parallaxes for $\log g$. The paper has a table with the solar abundances they employ. They use LTE 1D MARCS models and MOOG. Atomic data is provided as online material, line-by-line abundances are also provided in their online Table 11. The parameters  adopted for \hiprepVIII\ are:  $\mathrm{T}_{\mathrm{eff}} = 4920$ K, $\log g =2.30$, $v_{\mathrm{mic}} = 1.4$,  and $\mathrm{[Fe/H}  =-1.56$. The parameters for \hiprepX are:  $\mathrm{T}_{\mathrm{eff}} = 4500$ K, $\log g = 0.55$, $v_{\mathrm{mic}} = 1.95$,  and $\mathrm{[Fe/H}  = -2.93$. 

\subsection{R09 -  \cite{2009A&A...506..203R}}
This paper focuses on the analysis of \hiprepII, which was also targeted by CoRoT. The authors performed a spectroscopic analysis based on the ionization/excitation balance by looking at lines of HARPS spectra. The abundances are obtained using VALD atomic data and assuming LTE. Abundances are reported based on the \cite{grevesse07} solar scale. The parameters adopted in R09 for this star are around 100K cooler and have 0.2 dex lower surface gravity, though [Fe/H] agrees very well.

\subsection{S25 - \cite{2025A&A...701A.153S}}
This recent paper reports abundances of n-capture elements for about 150 stars in an attempt to investigate the chemical properties of planet hosts. Among these stars, \hiprepVII\ is included. Stellar parameters come from EWs measured with DAOSPEC  and MOOG, using the ionization/excitation balance approach. The abundances of the elements were determined by differential line-by-line spectrum synthesis using the TURBOSPECTRUM code with the MARCS stellar model atmospheres using the Sun as reference. When available, they perform non-LTE corrections using the TSFitPy wrapper \citep{2023A&A...669A..43G}. An extensive discussion on non-LTE corrections can be found that paper. On page 3, they list the lines used for the abundance analysis, as well as the adopted stellar parameters for this star ($T_{\mathrm{eff}} = 4862$ K, $\log g = 3.04$, and [Fe/H] = 0.04), which are quite different from the ones adopted in this work.

\subsection{S15 - \cite{2015A&A...584A..86S}}
This paper is dedicated to the abundance analysis of \hiprepIV\ observed with ESPaDOnS. The authors use ARES and TURBOSPECTRUM to determine abundances considering VALD and NIST linelist. The solar abundances come from \cite{2009Asplund}. While they perform non-LTE corrections using the code MULTI \citep{1999ARep...43..533K}, their Table 3 presents the LTE abundances, which we use to compare here. They also provide A(X) so we could transform these values to [X/Fe] using our solar reference from \cite{grevesse07} to match better our results. They have an extensive discussion of each element and which spectral lines they used for each element. ESPaDOnS has a more extended wavelength coverage than our library, which allows them to measure more abundances for \hiprepIV\ than us. 

\section{Line selection for clustering}\label{appendix:line_sel_clustering}
\begin{figure}[t]
    \centering
    \includegraphics[width=0.99\linewidth]{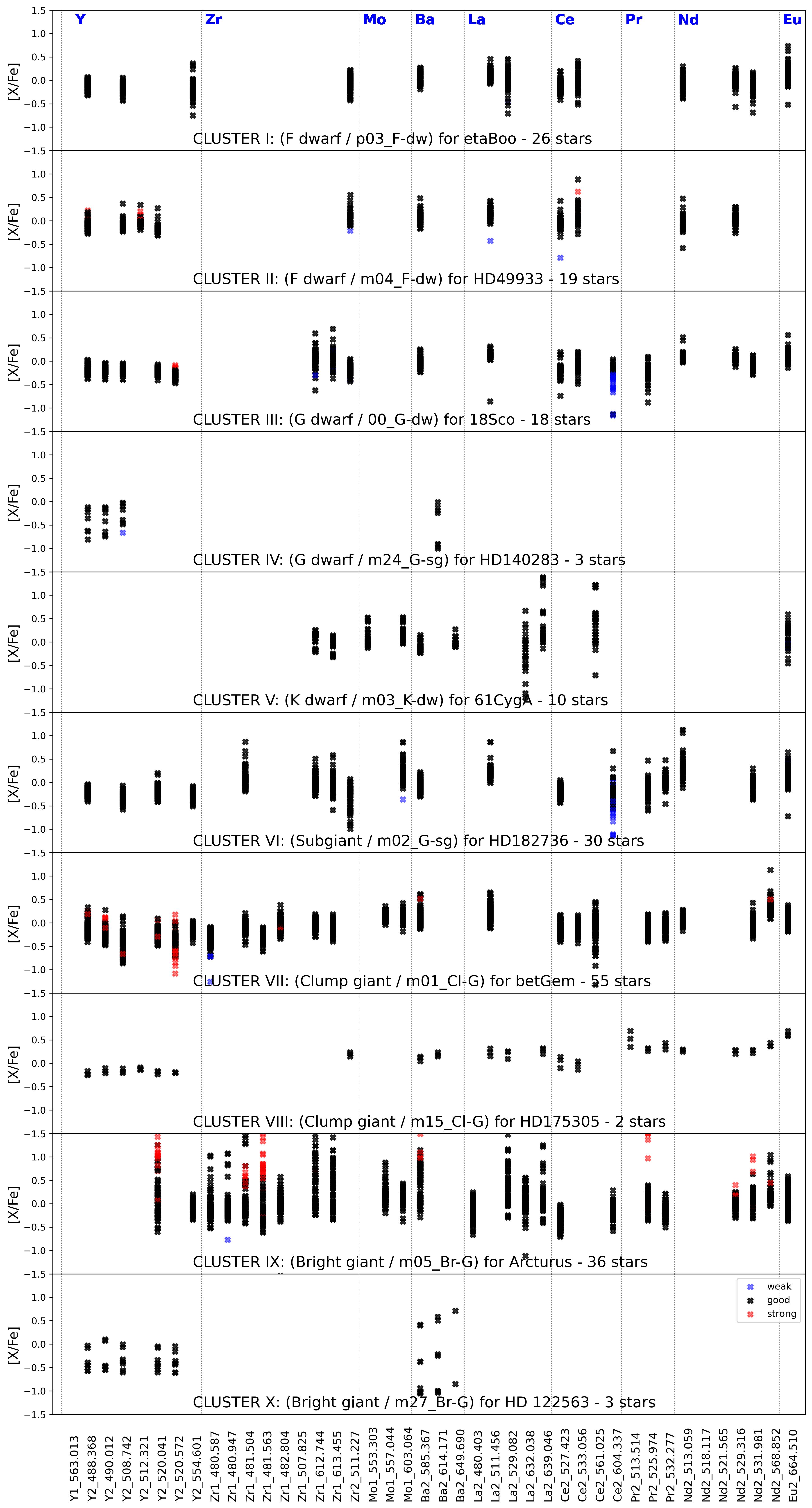}
    \caption{
    Similar to Fig.~\ref{fig:lines_new} but for all GBSv3 stars grouped according to the clustering algorithm and plotting only the selected lines of Table~\ref{tab:lines}. Abundances are shown in [X/Fe] form instead of absolute abundances and the groups and number of stars appear as title of each panel. Same color scheme as in Fig.~\ref{fig:lines_new}.}
    \label{fig:line_validation_clusters}
\end{figure}

Figure~\ref{fig:line_validation_clusters} is similar to Fig.~\ref{fig:lines_new} and shows the line selection for the clustering of the GBS sample. We adopt the same color scheme based on the REW criterion, namely black indicates that the REW falls within the desired range, while red indicates saturation. However, instead of the absolute abundance A(x), we use the abundance ratio relative to iron and solar values from \cite{grevesse07}. Only the lines selected in Table~\ref{tab:lines} are shown. Nevertheless, the abundance range for each line is larger here than in Fig.~\ref{fig:lines_new}, as values from all spectra of all stars within each group are included. The number of stars in each group is indicated in each panel, together with the name of the corresponding RepGBS.
\section{Solar abundances}

Table~\ref{Tab:solar} lists our solar abundances derived for the n-capture elements analyzed in this work together with their ionization stages (except for Zr, for which we adopted the average of \ion{Zr}{I} and \ion{Zr}{II}). The results are based on the line selection presented in Table~\ref{tab:lines}. For the Sun, we adopt the lines of the Group~III, that is, \classrepIII\ represented by \hiprepIII. Literature values are also reported for comparison.

\begin{table}[h!]
\caption{Comparison of solar abundances}
\centering
\begin{tabular}{|l|rrr|}
\hline
A(X) & GBS & GR07 & AS09 \\
\hline
\ion{Y}{II} & 1.98$\pm$0.04  & 2.21$\pm$0.02 & 2.21$\pm$0.05\\
Zr & 2.49$\pm$0.14 & 2.58$\pm$0.02 & 2.58$\pm$0.04\\
\ion{Ba}{II}& 2.13$\pm$0.03 & 2.17$\pm$0.07 & 2.18$\pm$ 0.09\\
\ion{La}{II}& 1.22$\pm$0.09 & 1.13$\pm$0.05 & 1.10$\pm$0.04 \\
\ion{Ce}{II}& 1.52$\pm$0.07 & 1.70$\pm$0.10 & 1.58$\pm$0.04\\
\ion{Pr}{II}& 0.49 $\pm$0.10 &   0.58 $\pm$0.10& 0.72$\pm$ 0.04\\
\ion{Nd}{II}& 1.46$\pm$0.09 &   1.45$\pm$0.05& 1.42$\pm$ 0.04\\
\ion{Eu}{II}& 0.57$\pm$0.05 & 0.52$\pm$0.06  & 0.52$\pm$ 0.04\\
\hline
\end{tabular}
\label{Tab:solar}
\tablefoot{The results derived in this work are listed in the GBS column, while the GR07 and AS09 columns report the results from \cite{grevesse07} and \cite{2009Asplund}, respectively.}
\end{table}

\section{Uncertainties}\label{app:errors}

\begin{figure}[t]
    \centering    \includegraphics[width=0.49\textwidth]{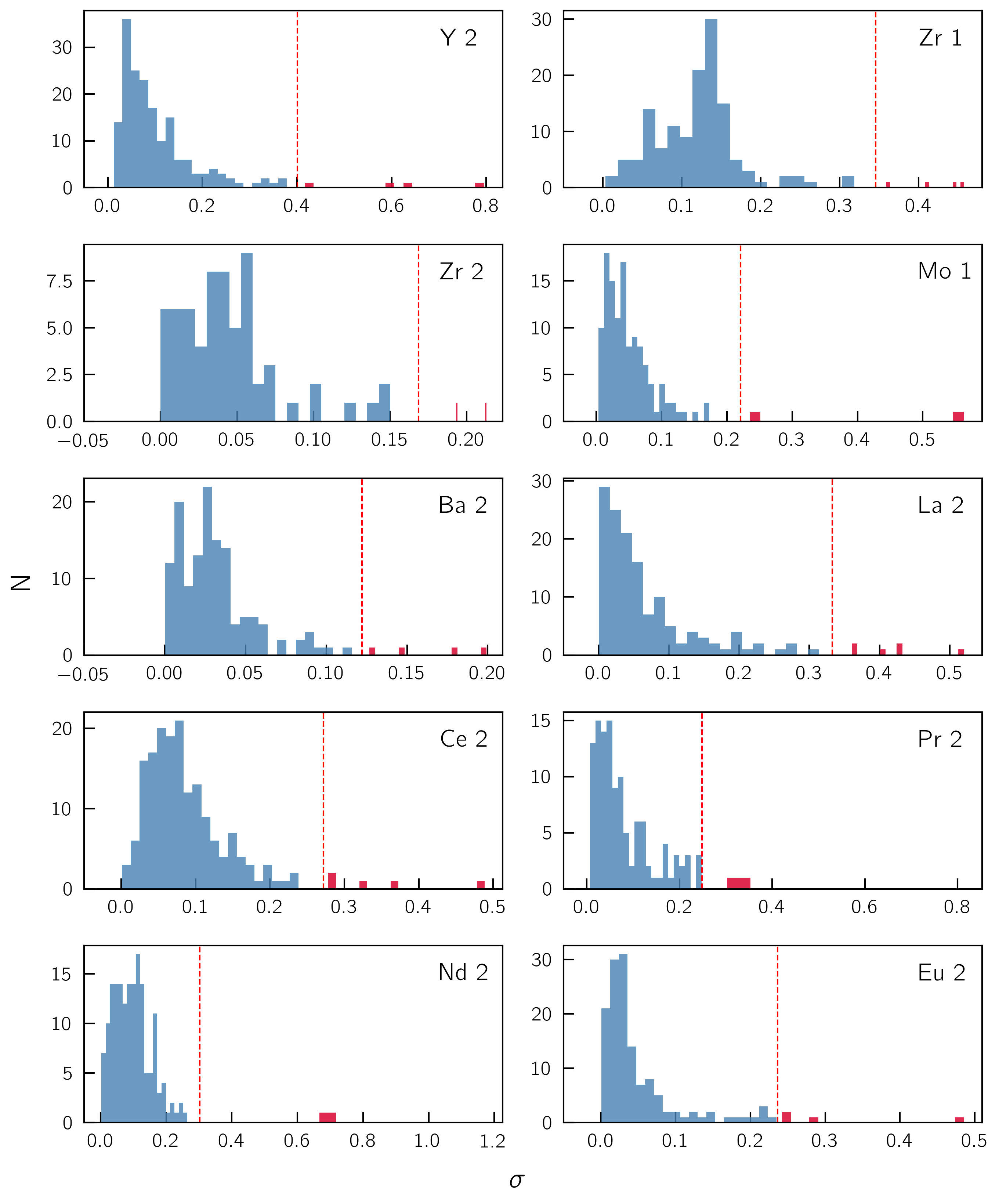}    
    \caption{Distribution of the weighted standard deviation ($\sigma$, from Eq.~\ref{uncert1}) for each element in the sample. Outliers are highlighted in red and excluded from the final sample (shown in blue) to ensure the robustness of the derived abundances. The vertical dashed line marks the threshold used to identify extreme values, defined as those beyond $3\sigma$. }\label{fig:hist_err}
    \label{fig:histo_err}
\end{figure}

In Fig.~\ref{fig:hist_err}, we report the uncertainties (computed with Eq.~\ref{uncert1}) for the elements derived in the final GBS sample. These distributions include only lines that pass the line selection, the REW cut, and the quality criteria on the synthetic line fits. The distributions lie mostly below $\sim 0.2$–$0.3$ dex, with a few outliers highlighted in red, corresponding to measurements beyond $3\sigma$ of the distribution. This additional filtering is applied to identify clear outliers that, despite passing all the quality cuts described in this work, may still affect the final abundances. Reassuringly, such cases are rare, with at most six measurements discarded for \ion{La}{II}. In this way, we further ensure the reliability of the final abundance determinations and validate our line-selection procedure. This results in the final number of measured elements per star, listed in Table~\ref{tab:element_counts}, which includes all 202 stars as well as the Sun.

\begin{table}[h!]
\caption{Number of stars with measured abundances for each n-capture element.}
\centering
\begin{tabular}{|l|rr|}
\hline
Element & $N_{\mathrm{stars}}$ & Percentage \\
\hline
\ion{Y}{II}  & 174 & 86.1\% \\
\ion{Zr}{I} & 135 & 66.8\% \\
\ion{Zr}{II} & 64  & 31.7\% \\
\ion{Mo}{I}  & 120 & 59.4\% \\
\ion{Ba}{II} & 133 & 65.8\% \\
\ion{La}{II} & 136 & 67.3\% \\
\ion{Ce}{II} & 168 & 83.2\% \\
\ion{Pr}{II} & 115 & 56.9\% \\
\ion{Nd}{II} & 165 & 81.7\% \\
\ion{Eu}{II} & 134 & 66.3\% \\
\hline
\end{tabular}
\label{tab:element_counts}
\end{table}

\end{document}